\documentclass{jfm}

\usepackage{amsmath,amssymb,graphics,overpic}

\usepackage{natbib}

\ifCUPmtlplainloaded \else
  \checkfont{eurm10}
  \iffontfound
    \IfFileExists{upmath.sty}
      {\typeout{^^JFound AMS Euler Roman fonts on the system,
                   using the 'upmath' package.^^J}%
       \usepackage{upmath}}
      {\typeout{^^JFound AMS Euler Roman fonts on the system, but you
                   dont seem to have the}%
       \typeout{'upmath' package installed. JFM.cls can take advantage
                 of these fonts,^^Jif you use 'upmath' package.^^J}%
      }
  \else
  \fi
\fi

\ifCUPmtlplainloaded \else
  \checkfont{msam10}
  \iffontfound
    \IfFileExists{amssymb.sty}
      {\typeout{^^JFound AMS Symbol fonts on the system, using the
                'amssymb' package.^^J}%
       \usepackage{amssymb}%

      }{}
  \fi
\fi

\ifCUPmtlplainloaded \else
  \IfFileExists{amsbsy.sty}
    {\typeout{^^JFound the 'amsbsy' package on the system, using it.^^J}%
     \usepackage{amsbsy}}
    {}
\fi

\DeclareMathOperator{\Bo}{B}
\DeclareMathOperator{\Ai}{Ai}
\DeclareMathOperator{\Bi}{Bi}

\title[Rivulet flow over a flexible beam]{Rivulet flow over a flexible beam}
\author[P. D. Howell, H. Kim, M. G. Popova and H. A. Stone]
{P. D. Howell$^1$%
\thanks{P. D. Howell and H. Kim contributed equally to this work},\ns H. Kim$^{2}\dagger$,\ns M. G. Popova$^{2}$ and
H. A. Stone$^{2}$\thanks{Email address for correspondence: hastone@princeton.edu}}
\affiliation{$^1$ Mathematical Institute, University of Oxford, Andrew Wiles Building, Oxford OX2 6GG, UK\\[\affilskip]
$^2$ Department of Mechanical and Aerospace Engineering, Princeton
University, Princeton, NJ 08544, USA}

\pubyear{????}
\volume{???}
\pagerange{???--???}
\date{?; revised ?; accepted ?. - To be entered by editorial office}

\begin{document}
\maketitle

\begin{abstract}
We study theoretically and experimentally how a thin layer of liquid
flows along a flexible beam.
The flow is modelled using lubrication theory and the substrate is
modelled as an elastica which deforms according to the Euler-Bernoulli
equation.
A constant flux of liquid is supplied at one end of the beam,
which is clamped horizontally, while the other end of the beam is
free.
As the liquid film spreads, its weight causes the beam deflection to
increase, which in turn enhances the spreading rate of the liquid.
This feedback mechanism
causes the front position $\sigma(t)$ and the deflection angle
at the front $\phi(t)$  to go through a number of different power-law
behaviours.
For early times, the liquid spreads like a horizontal gravity current,
with $\sigma(t)\propto t^{4/5}$ and $\phi(t)\propto t^{13/5}$.
For intermediate times, the deflection of the beam leads to rapid acceleration of the liquid layer, with $\sigma(t)\propto t^{4}$
and $\phi(t)\propto t^9$.
Finally, when the beam has sagged to become almost vertical, the liquid film flows downward with $\sigma(t)\propto t$ and $\phi(t)\sim\pi/2$.
We demonstrate good agreement between these theoretical predictions and experimental results.
\end{abstract}

\section{Introduction}

In the fluid mechanics literature, it is well known that similarity
solutions can describe the time-dependent spreading of thin viscous
films, which thus gives this nonlinear model problem great utility. A
similarly instructive problem from the elasticity literature concerns
the bending of a beam due to external forces and moments, which is
described by the Euler-Bernoulli equation and is nonlinear for large
changes in local orientation of the beam. It is then natural to couple
these two classical prototype problems from the mechanics literature
to consider how gravitational forces from a viscous film spreading
over a flexible beam can deflect the beam and so modify the shape and
propagation rate of the liquid film. We study this coupled
fluid-elastic dynamics problem using experiments and theory and
identify several distinct limits where there are similarity solutions
for the spreading rate and the beam deformation.

The general topic of elastohydrodynamics concerns problems where fluid flow is coupled to the deformation of an elastic boundary \citep{gohar2001elastohydrodynamics,dowson1999past}. Examples include the flow induced deformation of an elastic object or boundary during collision \citep{davis1986elastohydrodynamic}, droplet generation in a soft microfluidic device \citep{pang2014soft}, and the lift force on a sedimenting object generated by sliding motions accompanied by elastic deformation \citep{sekimoto1993mechanism,skotheim2005soft,salez2015elastohydrodynamics}. There are many natural examples related to a local flow-induced deformation, e.g. ejection of fungal spores from an ascus \citep{fritz2013natural}, biological tribology (articular cartilage) \citep{mow1992cartilage}, and raindrop impact on a leaf \citep{gart2015droplet,gilet2015fluid}. On the other hand, elastohydrodynamics also describes the movement of
a flexible solid object interacting with a surrounding flow, for example a micro-swimmer \citep{wiggins1998trapping,tony2006experimental}, an elastic fibre in a microchannel \citep{wexler2013bending}, or a flapping flag \citep{shelley2011flapping}.

Several previous studies have analysed the flow of a rivulet along a
prescribed inclined or curved substrate, for example 
\cite{Duffy1995141,EJM:43785,FLM:8825124,wilson2005uni}. 
Here our focus is a situation where the substrate geometry is unknown
in advance, and indeed is strongly coupled to the flow.
In our recent study \citep{howell2013gravity}, we developed a
two-dimensional model for steady gravity-driven thin film flow over a
flexible cantilever. In this paper, we analyse the flow of a liquid
rivulet along a flexible narrow beam, extending our previous study to
include time dependence and variations in the shape of the rivulet
cross-section.
We study theoretically and experimentally the time dependence of liquid propagation and beam deformation. The flow is modelled using lubrication theory and the substrate is modelled as an Euler-Bernoulli beam.
The related problem of flow of a layer of viscous fluid below an
elastic plate has been analysed for example by
\cite{flitton2004moving,lister2013viscous,hewitt2015elastic}, while
flow over an elastic membrane without bending stiffness was studied
theoretically and experimentally by \cite{zheng2015propagation}.

The paper is organised as follows. In \S 2 we present the experimental method and a large number of results for the beam deflection and rivulet propagation distance as functions of time. The experiments vary the bending modulus and length, width and thickness of the beam, and the flow rate of the liquid. In \S 3 we describe the governing equations and boundary conditions for the beam shape and the liquid film profile, demonstrating that the problems for the liquid spreading and the beam deformation are intimately coupled.
We find that the dynamics generically falls into one of two regimes,
namely a `small-deflection' regime and a `large-deflection' regime. We obtain similarity solutions to describe the time-dependent liquid propagation and the beam deflection for the different regimes.
We thus find three different power laws exhibited by the system during different time periods:
(i) at early times when the liquid just begins to deform the beam;
(ii) at intermediate times when the beam deflection increases rapidly in response to the weight of the liquid film;
(iii) at late times when the beam has sagged close to vertical.
We show that the experimental data collapse under scalings provided by the theoretical similarity solutions, and are then consistent with the theoretically predicted power laws.
Finally, we discuss the results and draw conclusions in \S\ref{sec:concs}.

\section{Experiments}

\subsection{Experimental setup}

\begin{figure}
  \centering
    \includegraphics[trim=0.0cm 0.0cm 0.0cm 0.0cm, clip=true, width=1\textwidth]{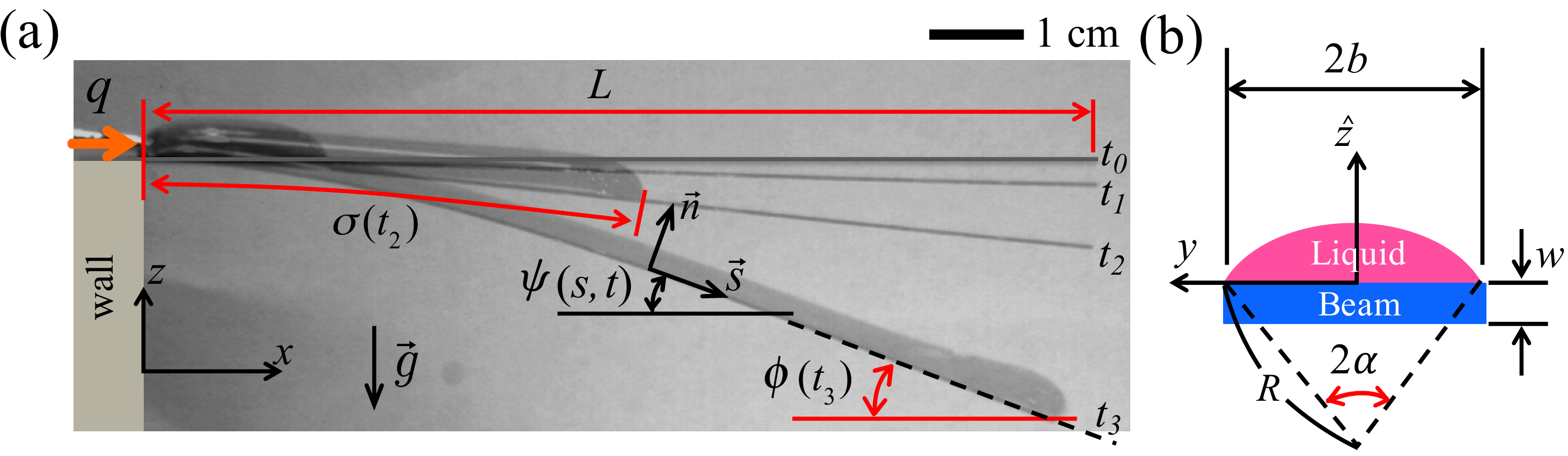}
  \caption{Experimental apparatus. (a) Side view: a thin elastic beam
    of length $L$ is fixed at the left wall and a constant flow rate
    $q$ is injected along the beam. The liquid wetted length is
    $\sigma(t)$ and the deflection angle at the advancing front is
    $\phi(t)$ where times $t_{0} < t_{1} < t_{2} < t_{3}$. Also,
    $\psi(s, t)$ is the local deformation angle, where $s$ is
    arc-length, while $\vec{s}$ and $\vec{n}$ are the unit tangent and
    normal vectors, respectively. (b) Front view: the cross-sectional
    shape of the liquid rivulet on the beam, where $2b$ and $w$
    denote, respectively, the beam width and thickness; $R$ is the
    radius of the curvature of the liquid-air interface and $2\alpha$
    is the opening angle. The $\hat{z}$-axis is in the direction of
    the normal $\vec{n}$.}
  \label{schematic}
\end{figure}

We performed experiments for liquid flow over a flexible cantilever. The experimental set-up is shown in figure~\ref{schematic}. The end of a thin elastic beam was fixed at a wall and a constant flow rate was applied by a syringe pump (Model: NE-1000, New Era Pump, USA).
In this study, we considered the effects of varying the flow rate $q$, as well as the Young's modulus $E$ of the beam, and the beam shape (i.e.\ length~$L$, width~$2b$, and thickness~$w$, as shown in figure~\ref{schematic}). For the liquid, we used glycerol (VWR International), which has dynamic viscosity
$\mu = 1.0\,\mathrm{Pa\,s}$,
density $\rho=1260\,\mathrm{kg/m}^{3}$, and surface tension $\gamma= 62.0\pm 0.5\,\mathrm{mN/m}$. To clearly observe the liquid propagation during the experiment, we added a red food dye (Innovating Science) to the liquid. The physical properties of the final liquid were measured at room temperature ($T = 298$\,K) with a rheometer (Anton-Paar MCR 301 with the CP 50 geometry) for the viscosity and with a conventional goniometer (Theta Lite, Biolin Scientific) for the surface tension.

Polycarbonate (PC) and polyether ether ketone (PEEK) were used as the material for the beam. To vary the bending stiffness,
we prepared various thicknesses
($w = 0.076$--$0.38$\,mm) and widths ($2b = 3$--$8$\,mm) of PC and PEEK materials (McMaster-Carr, NJ, USA). We obtained the Young's modulus of each material by measuring the self-deflection of the beam due to its own weight~\citep{crandall1978introduction}. The Young's moduli of PEEK and PC were measured as $E\approx 2.4$ and $3.5$\,GPa, respectively, which are consistent with the physical property values of the materials provided by the vendor. The two materials were initially covered by a protective film; before each experiment we removed the protective film and the beam was rinsed with distilled water and dried with nitrogen gas.

The deformation of the beam by the flowing liquid was observed from the side and top views, as shown in figure~\ref{expsetup}, using two CMOS color USB cameras (EO USB 2.0 with Nikon 1 V1 lens) with a frame rate of 1, 10, or 17 frames per second, and a spatial resolution of $1280\times1024$ pixels.
We measured the liquid propagation length $\sigma(t)$ and 
the deflection angle $\phi(t)$ at the advancing front,
as defined in figure~\ref{schematic}(a). To extract these quantities from the raw images, we performed image- and post-processing by using Matlab 2014a. We measured the evolution of $\sigma(t)$ and $\phi(t)$ up to the time when the liquid reached the end of the beam and began to drip.

\subsection{Experimental results}
\label{ss:expres}

\begin{figure}
  \centering
    \includegraphics[trim=0.0cm 0.0cm 0.0cm 0.0cm, clip=true, width=0.9\textwidth]{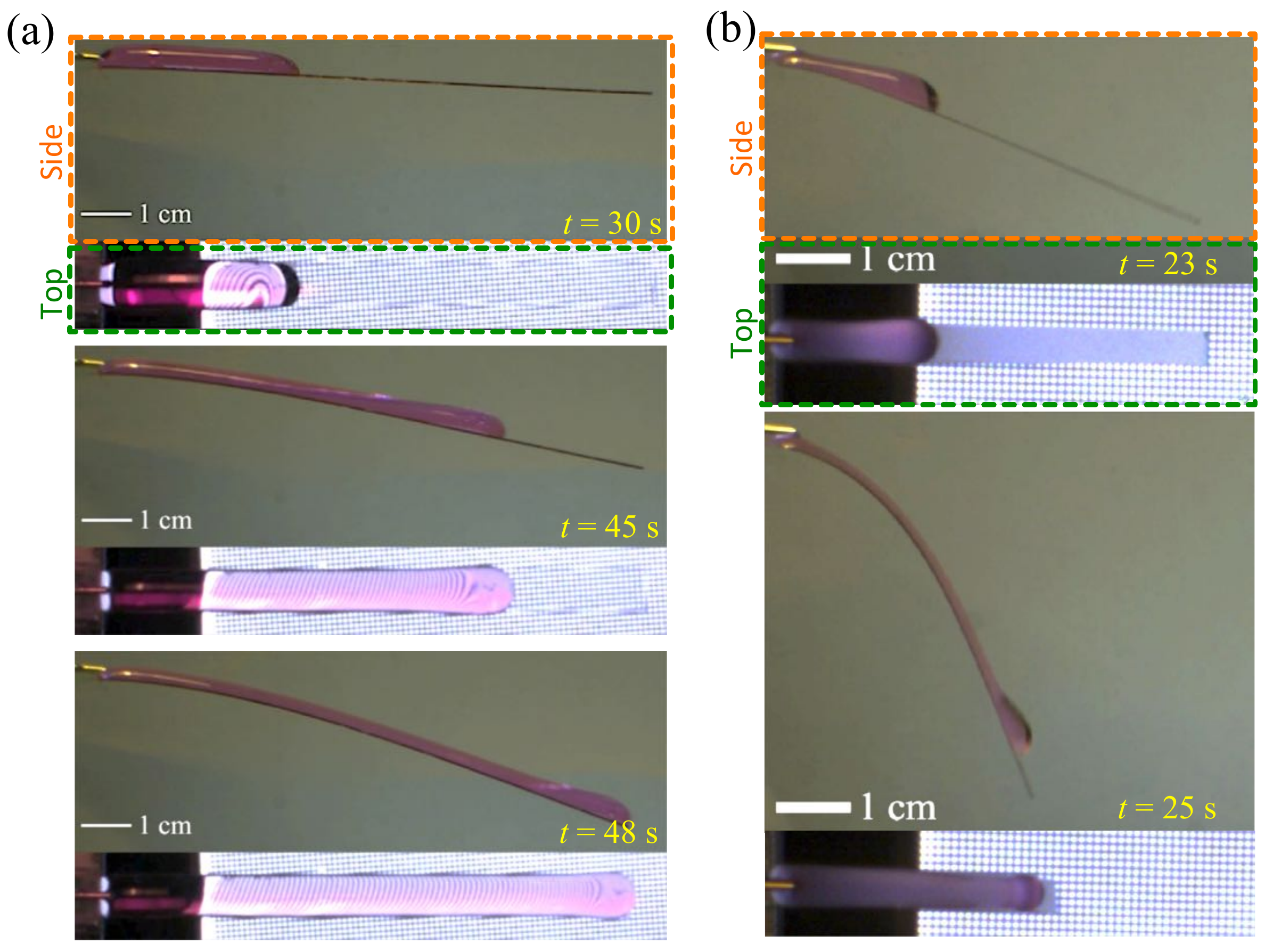}
  \caption{Examples of side and top views of liquid flow over an elastic beam. (a) A small beam deflection case with
$E=2.4$\,GPa, $q=1.4\times10^{-8}\,\mathrm{m^{3}/s}$,
$L=100$\,mm, $w=0.51$\,mm, and $2b=7$\,mm.
(b) A large beam deflection case with $E=3.6$\,GPa,
$q=2.2\times10^{-8}\,\mathrm{m^{3}/s}$,
$L=50$\,mm, $w=0.076$\,mm, and $2b=4$\,mm.}
  \label{expsetup}
\end{figure}

We investigate beam deformation and liquid propagation along the flexible beam while a constant flow rate is applied at the base.
Two typical examples of how the beam deformation and liquid film evolve over time are displayed 
in figure~\ref{expsetup} for two different values of the bending stiffness $Ebw^3/6$, namely
(a)~$1.84\times10^{-4}\,\mathrm{Pa\,m^{4}}$ and (b)~$5.31\times10^{-7}\,\mathrm{Pa\,m^{4}}$, respectively
(see also Supplementary Movie 1 and Supplementary Movie 2).
In case~(a), the relatively stiff beam suffers only a small deflection, such that the angle $\phi(t)<\pi/6$ up until the time when the liquid reaches the end of the beam; this is an example of what we refer to below as the ``small deflection'' regime.
Figure~\ref{expsetup}(b) shows the evolution of a much less stiff beam, which soon sags until the deflection angle $\phi(t)$ approaches $\pi/2$ and the liquid flow is close to vertical.
Below we refer to this more dramatic behaviour as the ``large deflection'' regime.

\begin{figure}
  \centering
    \includegraphics[trim=0.0cm 0.0cm 0.0cm 0.0cm, clip=true,
    width=1\textwidth]{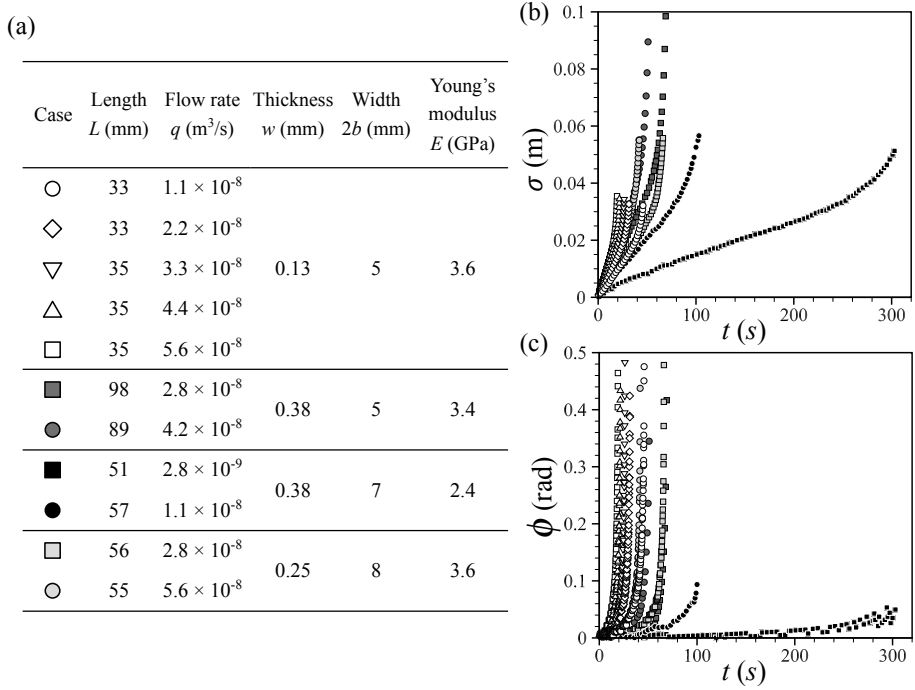}
  \caption{Small deflection results. (a)~Experimental parameters. (b)~Time evolution of the liquid propagation length $\sigma(t)$ (m). (c)~Time evolution of the deflection angle~$\phi(t)$ (rad) at the advancing front.}
  \label{small-results}
\end{figure}

For the small deflection regime, we summarise experimental conditions and results as shown in figure~\ref{small-results}. The flow rate~$q$, Young's modulus~$E$ and the beam dimensions $(L, 2b,$ and $w)$ are all varied, as listed in figure~\ref{small-results}(a), while the bending stiffness in each case is sufficient to keep the deflection angle less than $\pi/6$ throughout an experiment. Figures~\ref{small-results}(b) and~\ref{small-results}(c) show the time evolution of the liquid propagation length~$\sigma(t)$ and the deflection angle~$\phi(t)$. Initially, $\phi(t)$ remains close to zero, and the liquid spreads steadily, with $\sigma(t)$ apparently close to linear in $t$. However, the angle~$\phi(t)$ then increases rapidly, which in turn causes a rapid acceleration in the front position $\sigma(t)$.

\begin{figure}
  \centering 
    \includegraphics[trim=0.0cm 0.0cm 0.0cm 0.0cm, clip=true,
    width=1\textwidth]{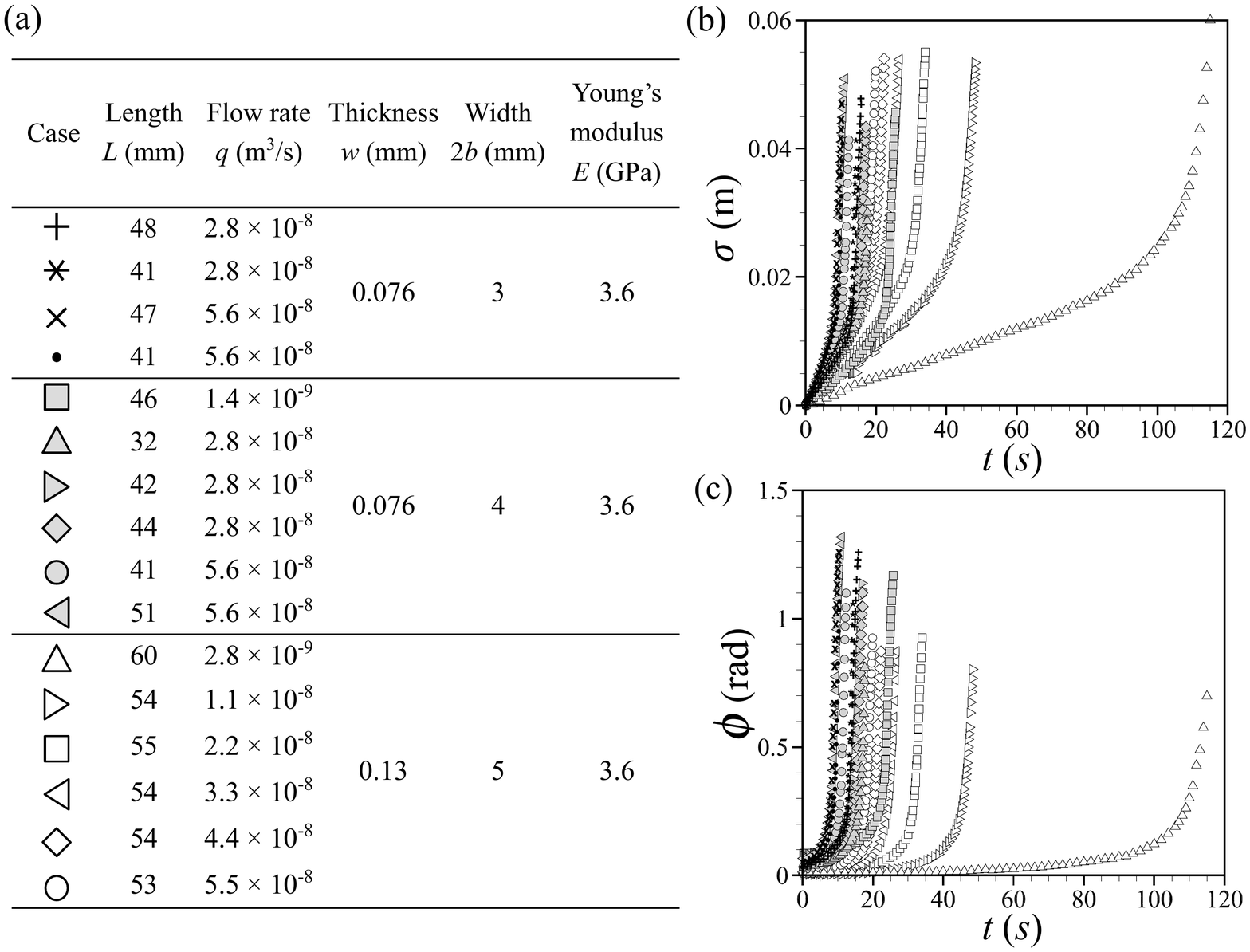}
  \caption{Large deflection results. (a)~Experimental parameters. (b)~Time evolution of the liquid propagation length~$\sigma(t)$ (m). (c)~Time evolution of the deflection angle $\phi(t)$ (rad) at the advancing front.}
  \label{large-results}
\end{figure}

Next, we present experimental results of the large deflection regime in figure~\ref{large-results}. The flow rate and beam geometry are again varied,
as shown in figure~\ref{large-results}(a), and
the corresponding time evolution of $\sigma(t)$ and $\phi(t)$
are shown in figures~\ref{large-results}(b) and~\ref{large-results}(c), respectively.
Compared with the results in figure~\ref{small-results}, the beams used here are thinner such that the deflection angle exceeds $\pi/6$ and, indeed, approaches $\pi/2$.
In some cases, the beam is initially slightly deformed by its weight, and there is also an angle measurement error of approximately 3$^{\circ}\approx 0.05$ radians. Thus, for some cases the beam deflection angle $\phi(t)$ appears to start from a non-zero value at $t=0$\,s. 

In the following section we present a theoretical model that describes the behaviour shown in figures~\ref{expsetup}--\ref{large-results} and allows the experimental results to be explained and quantitatively analysed.

\section{Mathematical theory}

\subsection{Governing equations}
\label{sec:geq}

We use Cartesian coordinates $(x,z)$ as shown in figure~\ref{schematic}, with the $z$-axis pointing vertically upwards and the beam clamped at $x=0$; the width of the beam lies in the $y$-direction.
We parametrise the deformation of the beam in the $(x,z)$-plane
using arc-length $s$ and time~$t$, such that
\begin{align}
\frac{\partial x}{\partial s}&=\cos\psi,
&
\frac{\partial z}{\partial s}&=-\sin\psi,
\end{align}
where $\psi(s,t)$ is the local angle made by the beam with the
$x$-axis (see the definitions in figure~\ref{schematic}(a)).

Let $A(s,t)$ denote the cross-sectional area of a thin liquid
film flowing over the top of the beam.
A one-dimensional mass conservation equation for the liquid is then
\begin{equation}\label{AtQs}
\frac{\partial A}{\partial t}+\frac{\partial Q}{\partial s}=0,
\end{equation}
where $Q(s,t)$ is the flux of liquid along the beam. We assume that a
known constant flux $q$ is supplied at the upstream end, so that
$Q(0,t)\equiv q$.

The tangential and normal components of the external
force per unit length exerted on
the beam are denoted by $f_s$ and $f_n$.
The Euler--Bernoulli equations governing the beam deformation are
then given by
\begin{align}\label{EulerBernoulli1}
\frac{\partial T}{\partial s}+N\frac{\partial\psi}{\partial s}&=
-f_s,
&
\frac{\partial N}{\partial s}-T\frac{\partial\psi}{\partial s}&=
-f_n,
&
EI\frac{\partial^2\psi}{\partial s^2}=N,
\end{align}
where $T$ and $N$ are the tension and shear force in the beam,
and
\begin{equation}
EI=\frac{Ebw^3}{6}
\end{equation}
is the bending stiffness.

To close the model, we need constitutive relations for the flux $Q$
and the components of the force/length $(f_s,f_n)$ in terms of $A$ and $\psi$.
Our aim in this study is to find a tractable model that adequately
captures the behaviour observed in experiments and is
amenable to mathematical analysis. To this end we make a number of
assumptions to obtain relatively simple closed-form constitutive
relations. First we neglect the contribution of the beam's own weight
to the stress components $f_s$ and $f_n$. In the experiments, the beam does sag somewhat by itself, e.g. see figure~\ref{large-results}(c), but this self-induced deflection is small
compared to the subsequent deflection once the fluid is injected, and
we have found that including the weight of the beam in the theory
makes very little difference to the results.
We thus obtain the following expressions for the components of the
force/length exerted on the beam by the fluid:
\begin{align}\label{ForcePerLength1}
f_s&=\rho gA\sin\psi-A\frac{\partial P}{\partial s},
&
f_n&=-\rho gA\cos\psi,
\end{align}
where $\rho$ is the density of the fluid, $g$ is the acceleration due
to gravity and $P(s,t)$ is the fluid pressure measured at the beam
surface.

Constitutive relations relating the pressure $P$ and flux $Q$ to
$A(s,t)$ and $\psi (s,t)$ may be formally derived using lubrication
theory in the limit where the fluid layer is relatively thin.
The simplified relations
\begin{align}\label{crPQsimp}
P(s,t)&=\left(\frac{3\gamma}{2b^3}\right)A,
&
Q(s,t)&=\frac{9A^3}{70\mu b^2}\left(\rho g\sin\psi-
\frac{3\gamma}{2b^3}\,\frac{\partial A}{\partial s}\right)
\end{align}
are derived in Appendix~\ref{ss:deriv} in the asymptotic limit where
the fluid layer is relatively thin and the
Bond number, $\Bo$, is small, i.e.
\begin{align}\label{assumption}
\frac{A}{b^2}&\ll1
& &\text{and}&
\Bo=\frac{\rho g b^2}{\gamma}&\ll1.
\end{align}
It must be acknowledged that neither of these assumptions holds
uniformly in the experiments. For example, based on the experimental
conditions, we estimated that $A/b^{2}\simeq 1$ and
$0.4\lesssim\Bo\lesssim3$. Nevertheless, we believe that the
approximations (\ref{crPQsimp}) are qualitatively reasonable and we
will use them henceforth.

Combining (\ref{AtQs}), (\ref{EulerBernoulli1}), (\ref{ForcePerLength1}), and (\ref{crPQsimp}), our final model equations are 
\begin{subequations}\label{dim:goveqs}
\begin{align}
\frac{\partial A}{\partial t}+\frac{9}{70\mu b^5}\,
\frac{\partial}{\partial s}\left[A^3\left(\rho gb^3\sin\psi  
-\frac{3\gamma}{2}\frac{\partial A}{\partial s}\right)\right]
&=0,
\\
\frac{\partial T}{\partial s}+N\frac{\partial\psi}{\partial s}
+\rho g A\sin\psi  
-\frac{3\gamma A}{2b^3}\frac{\partial A}{\partial s}&=0,
\\
\frac{\partial N}{\partial s}-T\frac{\partial\psi}{\partial s}-
\rho gA\cos\psi&=0,
\\ 
EI\frac{\partial^2\psi}{\partial s^2}&=N,
\end{align}
\end{subequations} 
which form a closed system for the four unknowns $A$, $\psi$, $T$ and $N$. 
The corresponding boundary conditions are 
\begin{subequations}
\begin{align}\label{dim:bcs0}
A^3\frac{\partial A}{\partial s}
+\frac{140\mu b^5 q}{27\gamma}
=\psi&=0 
&\text{at }s&=0,
\\ \label{dim:bcssi}
A=A^3\frac{\partial A}{\partial s}
=N=T=\frac{\partial\psi}{\partial s}
&=0 
&\text{at }s&=\sigma(t),
\end{align}
\end{subequations}
where $s=\sigma(t)$ denotes the moving front of the spreading rivulet. 
The conditions (\ref{dim:bcs0}) arise from the prescribed flux $q$ and horizontal clamping at $s=0$. 
The free boundary conditions (\ref{dim:bcssi}) arise from kinematic conditions for the liquid layer and from the imposition of no applied force or bending moment to the free end of the beam. 
The problem is closed by requiring the initial condition
$\sigma(0)=0$.

\subsection{Small deflection regime}
\label{ss:sdr}

\subsubsection{Normalised problem}

While the deflection angle $\psi$ is relatively small, the beam
equations may be linearised and the problem (\ref{dim:goveqs}) is then
approximated by
\begin{align}\label{sdr:dimeqs}
\frac{\partial A}{\partial t}+\frac{9}{70\mu b^5}\,
\frac{\partial}{\partial s}\left[A^3\left(\rho gb^3\psi  
-\frac{3\gamma}{2}\frac{\partial A}{\partial s}\right)\right]
&=0,
&
EI\frac{\partial^3\psi}{\partial s^3}&=\rho gA, 
\end{align}
where we have eliminated the force components $T$ and $N$.
The boundary conditions (\ref{dim:bcssi}) in terms of $A$ and $\psi$ are 
\begin{subequations}\label{sdr:dimbcs}
\begin{align}
A^3\frac{\partial A}{\partial s}
+\frac{140\mu b^5 q}{27\gamma}
=\psi&=0
&\text{at }s&=0,
\\ 
A=A^3\frac{\partial A}{\partial s}
=\frac{\partial\psi}{\partial s}=\frac{\partial^2\psi}{\partial s^2}
&=0 
&\text{at }s&=\sigma(t). 
\end{align}
\end{subequations}

\begin{figure}
\centering
\includegraphics[trim=0.0cm 0.0cm 0.0cm 0.0cm, clip=true, width=1\textwidth]{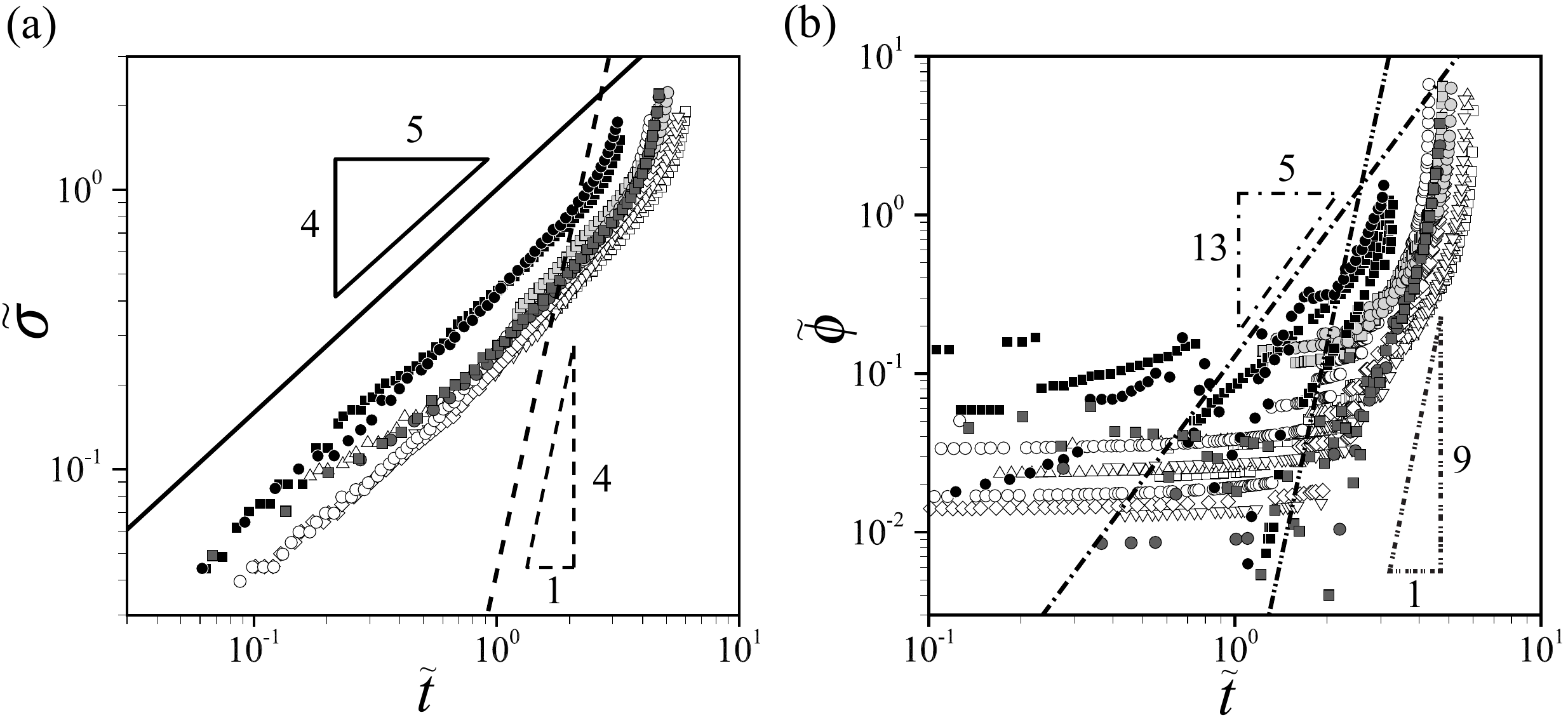}
\caption{Small deflection experimental results from
  figure~\ref{small-results} plotted using the
  normalised variables defined in (\ref{sdr:ndvars}):
(a)~$\tilde{\sigma}$ versus $\tilde{t}$ and (b)~$\tilde{\phi}$ versus
$\tilde{t}$. The asymptotic solutions in the small and large time limits are, respectively, 
$\tilde{\sigma} \sim 1.00101\tilde{t}^{4/5}$ (solid),
$\tilde{\sigma} \sim 0.0434638\tilde{t}^{4}$ (dash),
$\tilde{\phi} \sim 0.129117\tilde{t}^{13/5}$ (dash-dot), and
$\tilde{\phi} \sim 0.000302834\tilde{t}^{9}$ (dash-double-dot).}
\label{scale_results}
\end{figure}

The simplified problem (\ref{sdr:dimeqs})--(\ref{sdr:dimbcs}) may be
normalised by defining the dimensionless variables
\begin{subequations}\label{sdr:ndvars}
\begin{gather}
\tilde{A}=
\left(\frac{729}{9800}\right)^{1/8}
\left(\frac{\rho^{2}g^{2}\gamma^{3}}
{\mu^{4}q^{4}Eb^{18}w^{3}}\right)^{1/16}
A,
\\
\tilde{\psi}=
\left(\frac{2}{11025}\right)^{1/8}
\left(\frac{\rho^{10}g^{10}E^{3}b^{22}w^{9}}
{\mu^{4}q^{4}\gamma^{9}}\right)^{1/16}
\psi,
\\
\tilde{\sigma}=
\left(\frac{4\rho^{2}g^{2}b^{2}}
{\gamma E w^{3}}\right)^{1/4}
\sigma,
\qquad
\tilde{s}=
\left(\frac{4\rho^{2}g^{2}b^{2}}
{\gamma E w^{3}}\right)^{1/4}
s,
\\
\tilde{t}=
\left(\frac{1458}{1225}\right)^{1/8}
\left(\frac{\rho^{10}g^{10}q^{12}}
{\mu^{4}\gamma E^{5}b^{10}w^{15}}\right)^{1/16}
t.
\end{gather}
\end{subequations}
The rescaled variables satisfy the problem
(\ref{sdr:dimeqs})--(\ref{sdr:dimbcs}) with all the coefficients equal to unity, i.e.
\begin{subequations}\label{sdr:ndprob}
\begin{align}
\frac{\partial\tilde{A}}{\partial\tilde{t}}+
\frac{\partial}{\partial\tilde{s}}\left[{\tilde{A}}^3
\left(\tilde{\psi}-\frac{\partial\tilde{A}}{\partial\tilde{s}}
\right)\right]&=0,\label{FirstEqnDimensionlessA}
&
\frac{\partial^3\tilde{\psi}}{\partial\tilde{s}^3}&=\tilde{A},
\\
{\tilde{A}}^3\frac{\partial\tilde{A}}{\partial\tilde{s}}
+1=\tilde{\psi}&=0 
&\text{at }\tilde{s}&=0,
\\ 
\tilde{A}={\tilde{A}}^3\frac{\partial\tilde{A}}{\partial\tilde{s}}
=\frac{\partial\tilde{\psi}}{\partial\tilde{s}}
=\frac{\partial^2\tilde{\psi}}{\partial\tilde{s}^2}
&=0 &\text{at }\tilde{s}&=\tilde{\sigma}(t). 
\end{align}
\end{subequations}
In figure~\ref{scale_results}, we re-plot the small deflection experimental results
from figure~\ref{small-results} using the normalised variables
(\ref{sdr:ndvars}), and demonstrate that there is
indeed a reasonable collapse of the data.

\subsubsection{Small time limit}

As $\tilde{t}\rightarrow0$, we expect
$\tilde{\psi}\ll \bigl |{\partial\tilde{A}}/{\partial\tilde{s}}\bigr|$ 
in equation (\ref{FirstEqnDimensionlessA}).
In this limit, the problem becomes mathematically equivalent to a
classical gravity current on an effectively horizontal substrate
\citep{huppert1982propagation}. While a gravity current is driven by
hydrostatic pressure proportional to film height, in the present
problem, an analagous role is played by the capillary pressure
proportional to the cross-sectional area $A$.
The corresponding behaviour of the solution to the problem
(\ref{sdr:ndprob}) is described by a similarity solution of the form
\begin{align}
\tilde{A}\left(\tilde{s},\tilde{t}\right)&={\tilde{t}}^{1/5}f(\eta),
&
\tilde{\psi}\left(\tilde{s},\tilde{t}\right)&={\tilde{t}}^{13/5}g(\eta),
&
\eta&=\frac{\tilde{s}}{{\tilde{t}}^{4/5}},
\end{align}
where $f$ satisfies the ODE
\begin{align}\label{t0f}
f''+\frac{3(f')^2}{f}+\frac{4\eta f'}{5f^3}-\frac{1}{5f^2}&=0,
\end{align}
and the boundary conditions 
\begin{align}\label{t0bcs}
f^3(0)f'(0)&=-1,
& 
f(c)=\lim_{\eta\rightarrow c}f^3(\eta)f'(\eta)=0.
\end{align} 
The constant $c$ is to be determined as part of the solution, and the 
position of the free boundary is then given by
$\tilde{\sigma}\left(\tilde{t}\right)\sim c{\tilde{t}}^{4/5}$
as $\tilde{t}\rightarrow0$.
The deflection of the beam is determined
\textit{a~posteriori} from
\begin{align}\label{t0gint}
g(\eta)&=\frac{1}{2}\int_0^\eta f(\xi)\xi^2\,\mathrm{d}\xi
+\frac{1}{2}\int_\eta^c f(\xi)\eta(2\xi-\eta)\,\mathrm{d}\xi.
\end{align}

\begin{figure}
\resizebox{0.48\textwidth}{!}{\includegraphics{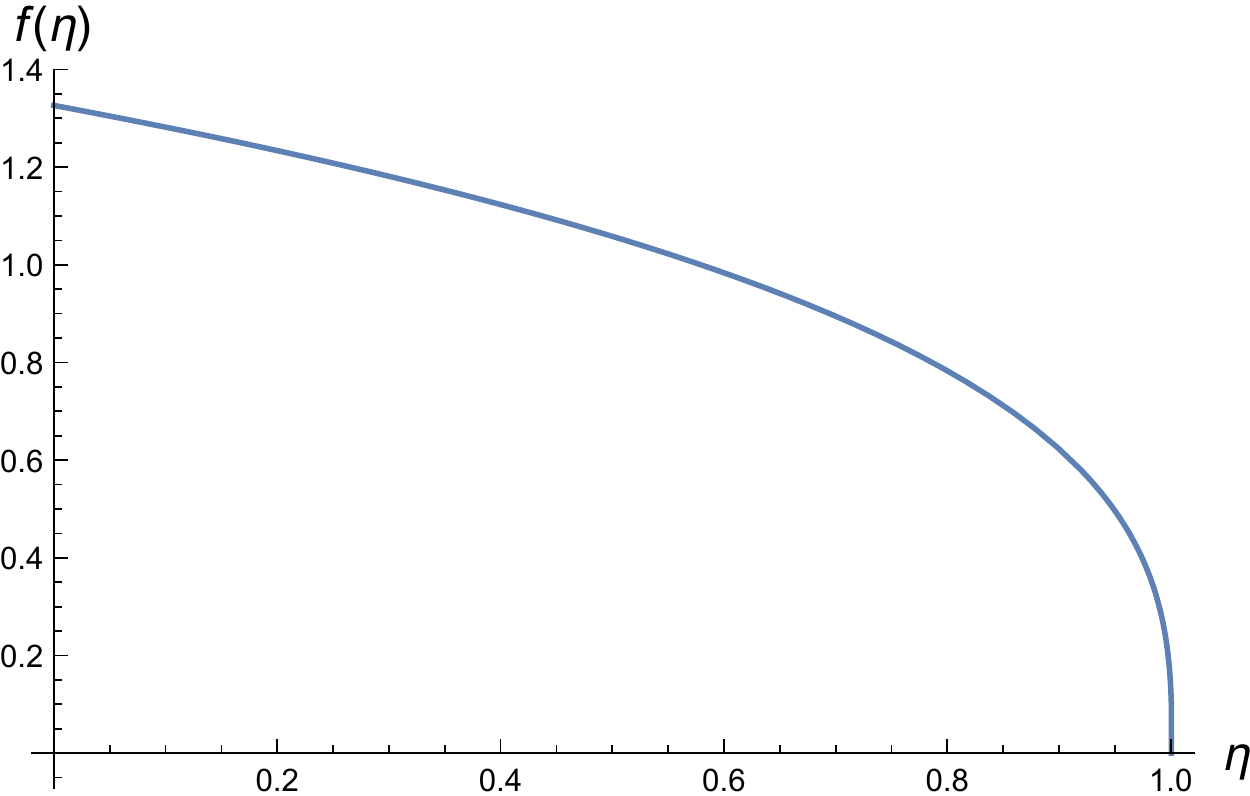}}%
\hfill%
\resizebox{0.48\textwidth}{!}{\includegraphics{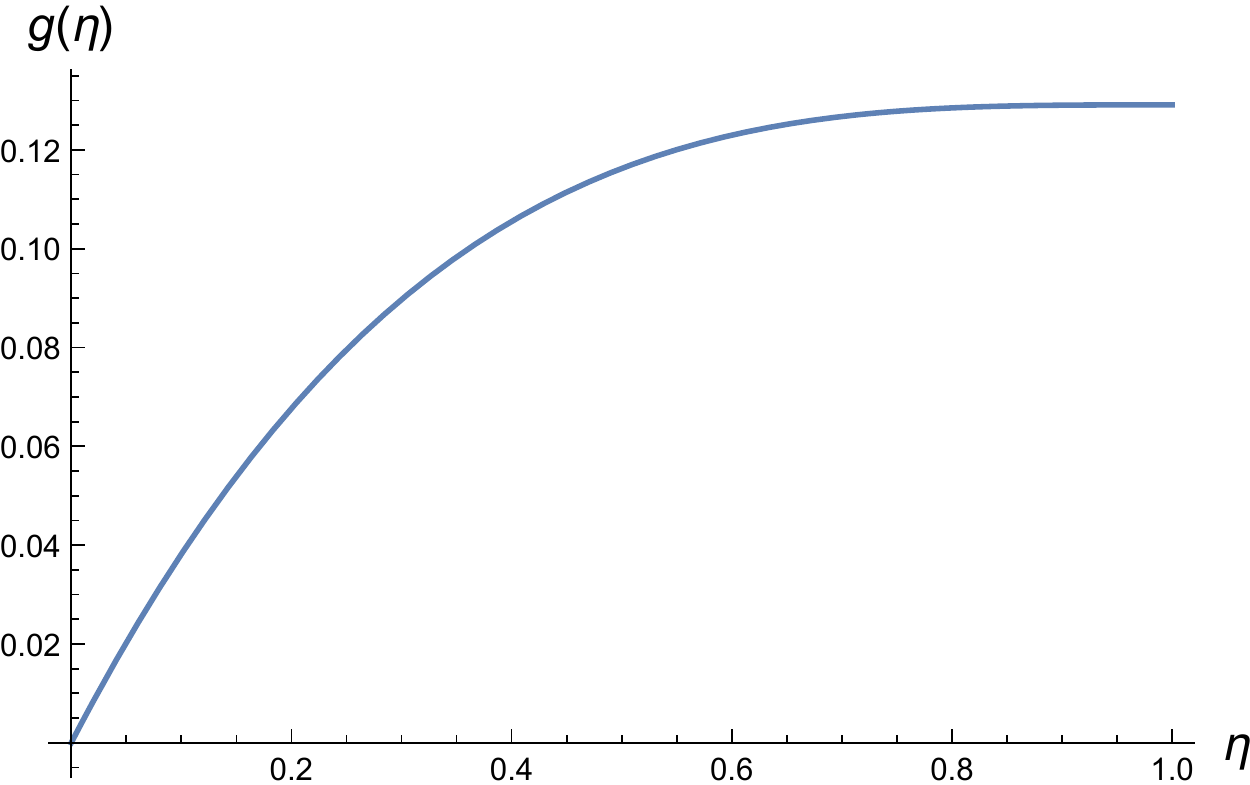}}
\caption{Small-$\tilde{t}$ similarity solution of the problem  
(\ref{t0f})--(\ref{t0gint}) for the normalised cross-sectional area  
$f(\eta)$ and deflection angle $g(\eta)$.}
\label{fig:smalltfg}
\end{figure}

The numerical shooting technique used to solve this problem is
outlined in Appendix~\ref{ss:sdst}, and the resulting solutions for
$f(\eta)$ and
$g(\eta)$ are plotted in figure~\ref{fig:smalltfg}.
The area profile resembles a classical gravity
current \citep{huppert1982flow,huppert1982propagation},
with a cube root singularity at the moving touch-down location $\eta=c$. 
From these solutions we read off the values
$c\approx1.00101$, $f(0)\approx1.32628$ and $g(c)\approx0.129117$.
Hence, in the small deflection regime, for small times the position of the advancing front and the maximum deflection angle at the front are given asymptotically by
\begin{align}\label{sdr:smalltsp}
\tilde{\sigma}\left(\tilde{t}\right)&\sim1.00101{\tilde{t}}^{4/5},
&
\tilde{\phi}\left(\tilde{t}\right)
=\tilde{\psi}\left(\tilde{\sigma}(\tilde{t}),\tilde{t}\right)
&\sim0.129117{\tilde{t}}^{13/5}
&&\text{as }\tilde{t}\rightarrow0.
\end{align}

The predicted power laws (\ref{sdr:smalltsp}) for
$\tilde{\sigma}\left(\tilde{t}\right)$ and
$\tilde{\phi}\left(\tilde{t}\right)$ are shown in
figure~\ref{scale_results}, using solid and dash-dotted lines,
respectively. There appears to be a good fit for the behaviour of
$\tilde{\sigma}$, so long as the deflection angle remains small.
The fit for $\tilde{\phi}$ is also quite good for a range of
intermediate times. The significant departures observed at very small
values of $\tilde{t}$ are due to the small initial deflection of the
beam under its own weight, which is not included in our model,
as well as angle measurement errors, as explained in
\S\ref{ss:expres}.

\subsubsection{Large time limit}

The limiting behaviour (\ref{sdr:smalltsp}) describes the evolution while the beam deflection remains small enough to have a negligible influence on the spreading of the liquid. As $\tilde{t}$ increases, the coupling between liquid flow and beam deformation becomes important. Eventually,
as $\tilde{t}\rightarrow\infty$, the non-dimensional flux term in square brackets in equation (\ref{sdr:ndprob}) is dominated by ${\tilde{A}}^3\tilde{\psi}$. In this case the limiting
behaviour is described by a similarity solution of the form
\begin{align}\label{sdr:ltss}
\tilde{A}\left(\tilde{s},\tilde{t}\right)&={\tilde{t}}^{-3}f(\eta),
&
\tilde{\psi}\left(\tilde{s},\tilde{t}\right)&={\tilde{t}}^9g(\eta),
&
\eta&=\frac{\tilde{s}}{{\tilde{t}}^4},
\end{align} 
where $f$ and $g$ satisfy the ODEs
\refstepcounter{equation}\label{tinffgde}
\begin{align*}
\left(f^3g-4\eta f\right)'+f&=0,
&
g'''-f&=0.
\tag{\theequation \textit{a,b}}
\end{align*}
The corresponding boundary conditions, including the imposed flux, are
\begin{align}\label{tinffgbcs}
g(\eta)\rightarrow0,\quad f(\eta)^3g(\eta)\rightarrow1
\quad\text{as }\eta\rightarrow0,
&&&
g'(c)=g''(c)=0.
\end{align}
Again the constant $c$ is to be determined as part of the solution,
and the large-$\tilde{t}$ behaviour of the free boundary is then given 
by $\tilde{\sigma}\left(\tilde{t}\right)\sim c\tilde{t}^4$. 
To close the problem, we note that a constant liquid flux imposes the net conservation equation
\begin{equation}
\int_0^c f(\eta)\,\mathrm{d}\eta=1.
\end{equation}
By integrating equation (\ref{tinffgde}\textit{a}) with respect to $\eta$, this integral condition may equivalently be stated as the boundary condition
\begin{equation}\label{tinfflux}
f(c)^2g(c)=4c.
\end{equation}

\begin{figure}
\resizebox{0.48\textwidth}{!}{\includegraphics{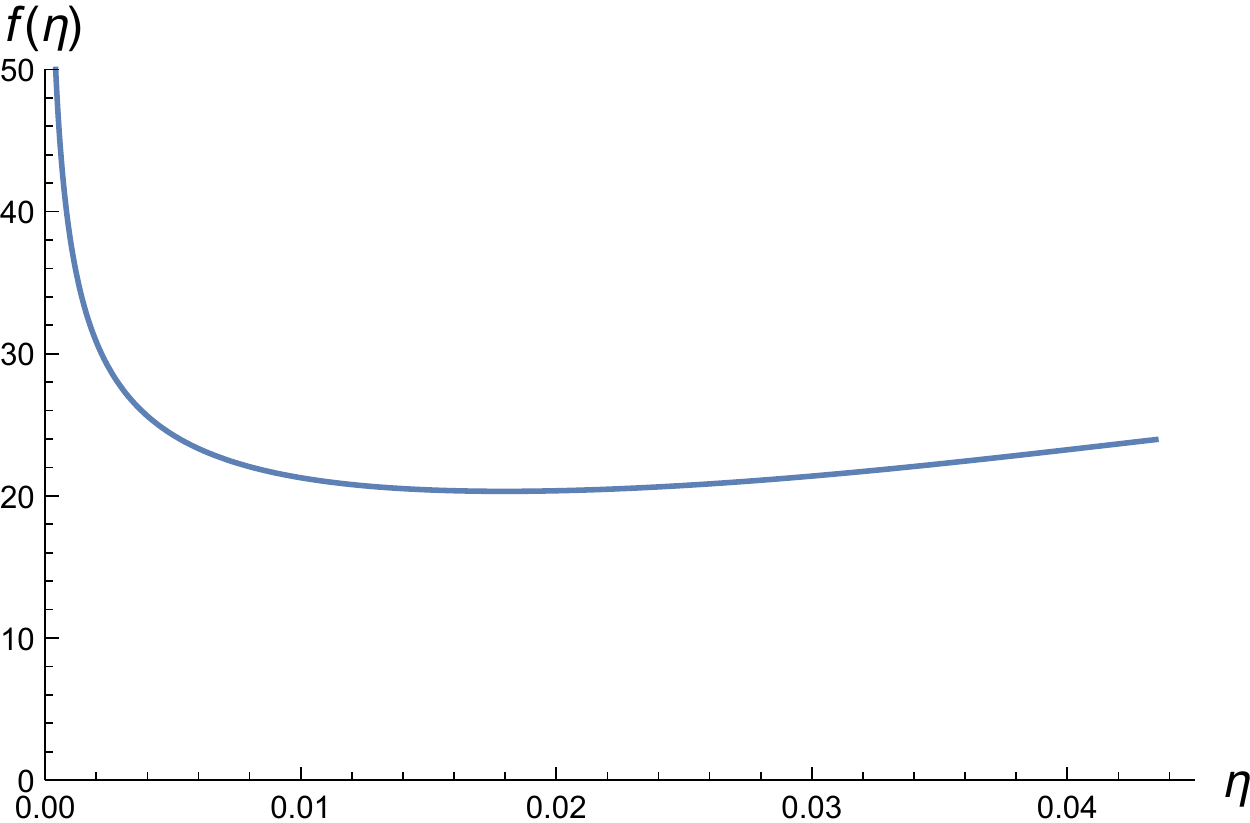}}%
\hfill
\resizebox{0.48\textwidth}{!}{\includegraphics{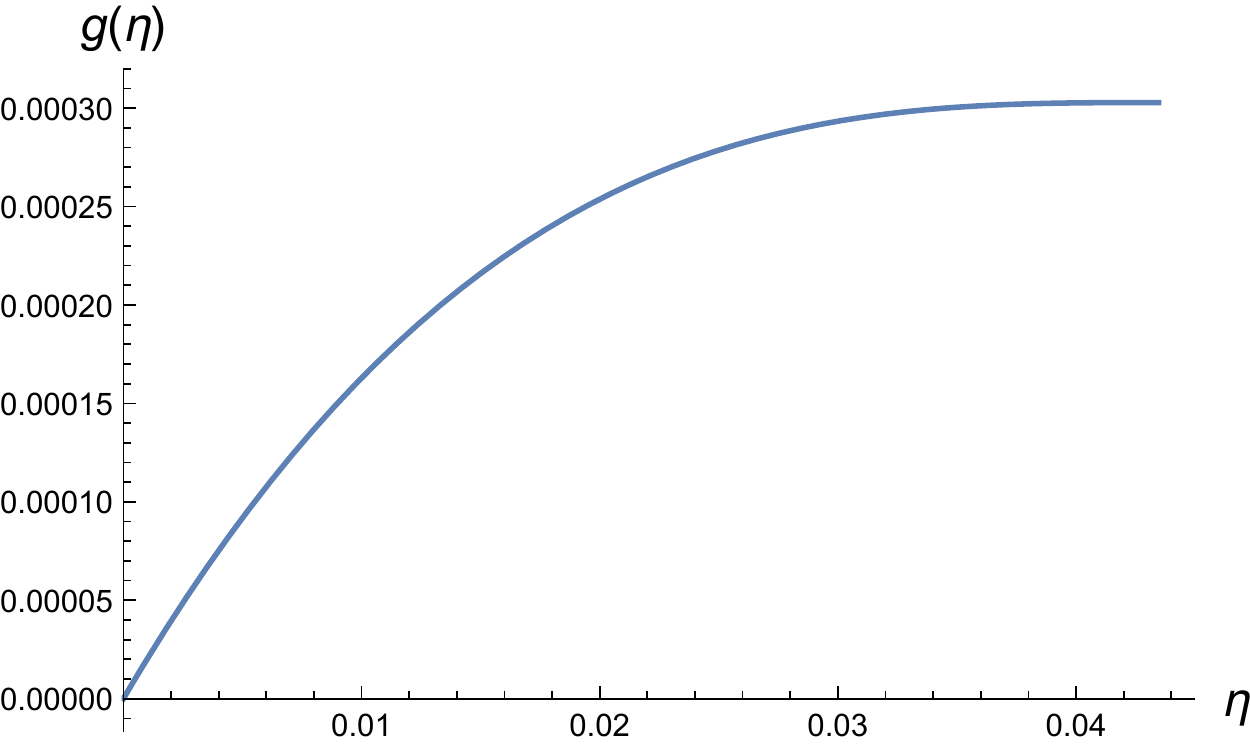}}
\caption{Large-$t$ similarity solution of the problem 
 (\ref{tinffgde})--(\ref{tinfflux}) 
  for the normalised cross-sectional area $f(\eta)$
and the normalised deflection angle $g(\eta)$.}
\label{fig:fgvseta}
\end{figure}

The boundary-value problem (\ref{tinffgde})--(\ref{tinfflux})
is solved using a shooting method outlined
in Appendix~\ref{ss:sdlt}, and the resulting solutions for $f(\eta)$
and $g(\eta)$ are plotted in figure~\ref{fig:fgvseta}.
We note that $f(\eta)$ decreases as $\eta$ increases from zero,
attains a minimum value of approximately $20.3181$ at 
$\eta\approx0.0179634$, and then 
increases again as $\eta$ approaches $c$.
This behaviour reflects well 
the non-monotonic profiles for the film thickness observed in the experimental results, as shown
in figure~\ref{expsetup}.
However, the problem
(\ref{tinffgde})--(\ref{tinfflux}) predicts that
$f(\eta)\sim3.64271\eta^{-1/3}$ as $\eta\rightarrow0$,
implying that the cross-sectional area diverges toward the origin;
also, we are unable to impose the condition $f(c)=0$ corresponding to 
the condition $\tilde{A}=0$ at the advancing front. 
Both of these apparent difficulties can be resolved by analysing 
asymptotic boundary layers near $\tilde{s}=0$ and 
$\tilde{s}=\tilde{\sigma}\left(\tilde{t}\right)$, as demonstrated 
in \cite{howell2013gravity} for the steady version of the problem. 

From the numerical solutions plotted in figure~\ref{fig:fgvseta}, we read off the values
\mbox{$c\approx0.0434638$},
\mbox{$f(c)\approx23.9603$},
\mbox{$g'(0)\approx0.0206883$},
and
\mbox{$g(c)\approx0.000302834$}.
Hence, in the small deflection regime, for large times the position of
the advancing front and the maximum
deflection angle are given asymptotically by
\begin{align}\label{sdr:siptinf}
\tilde{\sigma}\left(\tilde{t}\right)&\sim0.0434638{\tilde{t}}^4,
&
\tilde{\phi}\left(\tilde{t}\right) 
&\sim0.000302834{\tilde{t}}^9
&&\text{as }\tilde{t}\rightarrow\infty.
\end{align}
The power laws predicted in equation (\ref{sdr:siptinf}) are shown in 
figure~\ref{scale_results}, using dashed and dash-double-dotted lines,
respectively.
We observe that these power laws do give a reasonable fit to the
dramatic increase in the deflection angle and consequent
rapid movement of the rivulet along the beam.

\subsection{Large deflection regime}

\subsubsection{Normalised problem}

The power laws (\ref{sdr:siptinf}) are valid in an intermediate regime
where there is significant feedback between the beam deflection and
the liquid flow, but the deflection angle remains relatively small.
However, if the beam is sufficiently long, then
the assumption that $\psi\ll1$ must eventually fail, so that the
nonlinear terms in $\psi$ that were neglected in the linearised
problem (\ref{sdr:dimeqs}) become significant.
However, when $\psi=O(1)$, the capillary terms involving spatial
derivatives of $A$ in the governing equations (\ref{dim:goveqs})
become negligible compared with the gravitational terms
\citep[see][]{howell2013gravity},
and the equations may be simplified to
\begin{subequations}\label{ldr:dimeqs}
\begin{align}\label{ldr:dimAt}
\frac{\partial A}{\partial t}+\frac{9 \rho g}{70\mu b^2}\,
\frac{\partial}{\partial s}\left(A^3\sin\psi\right)
&=0,
\\\label{ldr:dimTs}
\frac{\partial T}{\partial s}+N\frac{\partial\psi}{\partial s}
+\rho g A\sin\psi&=0,
\\\label{ldr:dimNs}
\frac{\partial N}{\partial s}-T\frac{\partial\psi}{\partial s}-
\rho gA\cos\psi&=0,
\\ \label{ldr:dimPsi}
EI\frac{\partial^2\psi}{\partial s^2}&=N.
\end{align}
\end{subequations}

As in \cite{howell2013gravity}, a first integral of
(\ref{ldr:dimTs})--(\ref{ldr:dimNs}) allows us to write 
\begin{align}
T&=F\sin\psi,
&
N&=-F\cos\psi,
\end{align}
where $F$ is the vertical component of stress in the beam. 
The leading-order large-deflection equations
(\ref{ldr:dimTs})--(\ref{ldr:dimPsi}) therefore reduce to  
\begin{align}\label{ldr:dimFs}
\frac{\partial F}{\partial s}&=-\rho g A,
&
EI\frac{\partial^2\psi}{\partial s^2}&=-F\cos\psi,
\end{align}
which, with (\ref{ldr:dimAt}), form a closed system for $A$, $\psi$
and $F$.
The boundary conditions for $\psi$ and $F$ are
\begin{align}\label{ldr:dimpFbcs}
\psi=0 \quad\text{at }s&=0,
&
F=\frac{\partial\psi}{\partial s}=0\quad \text{at }s&=\sigma(t),
\end{align}
corresponding to horizontal clamping at $s=0$ and zero applied force
and moment at the free end of the beam.

Now that the highest spatial derivatives of $A$ have been neglected,
it is impossible to satisfy exactly the boundary conditions for $A$ at
$s=0$ and $s=\sigma(t)$. Instead, we impose the net flux conditions
\begin{align}\label{ldr:dimAbcs}
\frac{9\rho g A^3\sin\psi}{70\mu b^2}\rightarrow q 
\quad\text{as }s&\rightarrow0,
&
\frac{\mathrm{d}\sigma}{\mathrm{d}t}=
\frac{9\rho g A^2\sin\psi}{70\mu b^2}
\quad\text{at }s&=\sigma(t).
\end{align}
The full boundary conditions for $A$ may be imposed by analysing
asymptotic boundary layers near $s=0$ and $s=\sigma(t)$, in which the
spatial derivatives of $A$ regain their significance, as shown in
\cite{howell2013gravity}.

Now the problem (\ref{ldr:dimAt}),
(\ref{ldr:dimFs})--(\ref{ldr:dimAbcs})
may be normalised by introducing the new dimensionless variables
\begin{subequations}\label{ldr:nondim}
\begin{align}
\hat{A}&=\left(\frac{9\rho g}{70\mu qb^2}\right)^{1/3}A,
&
\hat{\sigma}&=
\left(\frac{1680\rho^2g^2\mu q}{w^9bE^3}\right)^{1/9}\sigma,
\\
\hat{s}&=
\left(\frac{1680\rho^2g^2\mu q}{w^9bE^3}\right)^{1/9}s,
&
\hat{t}&=\left(\frac{4374\rho^5g^5q^7}{1225\mu^2w^9b^7E^3}\right)^{1/9}t,
\end{align}
\end{subequations}
with respect to which the governing equations read
\refstepcounter{equation}\label{ldr:ndeqs}
\begin{align*}
\frac{\partial\hat{A}}{\partial\hat{t}}+
\frac{\partial}{\partial\hat{s}}\left({\hat{A}}^3\sin\psi\right) 
&=0,
&
\frac{\partial^2\psi}{\partial{\hat{s}}^2}+
\cos\psi\int_{\hat{s}}^{\hat{\sigma}}\hat{A}\left(s',t\right)\,\mathrm{d}s'
&=0
\tag{\theequation \textit{a,b}}
\end{align*}
subject to
\begin{subequations}\label{ldr:ndbcs}
\begin{align}
\psi\rightarrow0,~~{\hat{A}}^3\sin\psi\rightarrow1
\quad\text{as }\hat{s}&\rightarrow0,
\\
\frac{\partial\psi}{\partial\hat{s}}
=\frac{\mathrm{d}\hat{\sigma}}{\mathrm{d}\hat{t}}-
{\hat{A}}^2\sin\psi  
=0\quad\text{at }\hat{s}&=\hat{\sigma}\left(\hat{t}\right).  
\end{align}
\end{subequations}
Thus, once the fluid layer has progressed so far along the beam that
the deflection angle $\psi$ is $O(1)$, we expect the new scalings
(\ref{ldr:nondim}) to collapse the experimental data: this prediction
will be confirmed below.

As $\hat{t}\rightarrow0$, the solution of the problem
(\ref{ldr:ndeqs})--(\ref{ldr:ndbcs}) may be described by a similarity
solution that is equivalent to the large-$\tilde{t}$ solution
(\ref{sdr:ltss}). This result just confirms that the small- and
large-deflection regimes are mutually consistent for intermediate
values of~$\psi$.

\subsubsection{Large time limit}

At large values of $\hat{t}$, assuming the beam is sufficiently long, the weight of the fluid causes the beam to sag until it is approximately vertical.
To study this limit, we write $\psi=\pi/2-\chi$ where $0<\chi\ll1$:
it will transpire that $\chi$ is exponentially small.
The governing equation (\ref{ldr:ndeqs}\textit{a}) for $\hat{A}$ thus becomes
\begin{subequations}\label{ldr:thinf}
\begin{equation}
\frac{\partial\hat{A}}{\partial\hat{t}}+
\frac{\partial}{\partial\hat{s}}\left({\hat{A}}^3\right)
=0,
\end{equation}
which is subject to
\begin{align}
{\hat{A}}^3\rightarrow1\quad
\text{as }\hat{s}&\rightarrow0,
&
\frac{\mathrm{d}\hat{\sigma}}{\mathrm{d}\hat{t}}=
{\hat{A}}^2
\quad \text{at }\hat{s}&=\hat{\sigma}\left(\hat{t}\right).
\end{align}
\end{subequations}
The relevant large-$\hat{t}$ limiting solution of the problem
(\ref{ldr:thinf}) is
\begin{align}\label{ldr:nvAsi}
\hat{A}\left(\hat{s},\hat{t}\right)&=1,
&
\hat{\sigma}\left(\hat{t}\right)&=\hat{t}.
\end{align}

With $\hat{A}$ given by (\ref{ldr:nvAsi}), the
deflection equation (\ref{ldr:ndeqs}\textit{b}) reduces to
a form of the Airy equation for $\chi$:
\begin{equation}\label{ldr:Airy}
\frac{\partial^2\chi}{\partial{\hat{s}}^2}=
\left(\hat{\sigma}-\hat{s}\right)\chi.
\end{equation}
Given $\partial\chi/\partial\hat{s}=0$ at $\hat{s}=\hat{\sigma}$, the
solution of equation (\ref{ldr:Airy}) is
\begin{equation}\label{ldr:chisol}
\chi\left(\hat{s},\hat{t}\right)=\frac{3^{1/6}\Gamma(2/3)C}{2}
\bigl[\Bi\left(\hat{\sigma}-\hat{s}\right)
+\sqrt{3}\Ai\left(\hat{\sigma}-\hat{s}\right)\bigr],
\end{equation}
where $\Ai$ and $\Bi$ denote Airy functions and
$C(t)=\chi\left(\hat{\sigma},\hat{t}\right)$ is an arbitrary
integration function, equal to the value of $\chi$ at the advancing
front.

To determine $C$, and thus the deviation of the deflection from
vertical, we have to match with an inner region near $\hat{s}=0$ in
which $\psi$ rapidly adjusts from $0$ to almost $\pi/2$.
In this region, to lowest order the deflection equation
(\ref{ldr:ndeqs}\textit{b}) reduces to
\begin{equation}\label{ldr:inpss}
\frac{\partial^2\psi}{\partial\hat{s}^2}+\hat{\sigma}\cos\psi=0.
\end{equation}
The solution of (\ref{ldr:inpss}) subject to the boundary and matching
conditions
\begin{align}
\psi=0\quad\text{at }\hat{s}=0,
&&&
\psi\rightarrow\pi/2\quad\text{as }\hat{s}\rightarrow\infty
\end{align}
is
\begin{equation}\label{ldr:pinner}
\psi\left (\hat{s}, \hat{t}\right )=\frac{\pi}{2}-4\tan^{-1}
\left((\sqrt{2}-1)\mathrm{e}^{-\hat{s}\sqrt{\hat{\sigma}}}\right).
\end{equation}
Finally, we get an expression for $C$ by
matching (\ref{ldr:pinner}) with (\ref{ldr:chisol}):
\begin{equation}
C=\frac{8\sqrt{\pi}\left(\sqrt{2}-1\right) 
{\hat{\sigma}}^{1/4}\mathrm{e}^{-2{\hat{\sigma}}^{3/2}/3}}
{3^{1/6}\Gamma(2/3)}.
\end{equation}

\begin{figure}
  \centering 
    \includegraphics[trim=0.0cm 0.0cm 0.0cm 0.0cm, clip=true,
    width=0.8\textwidth]{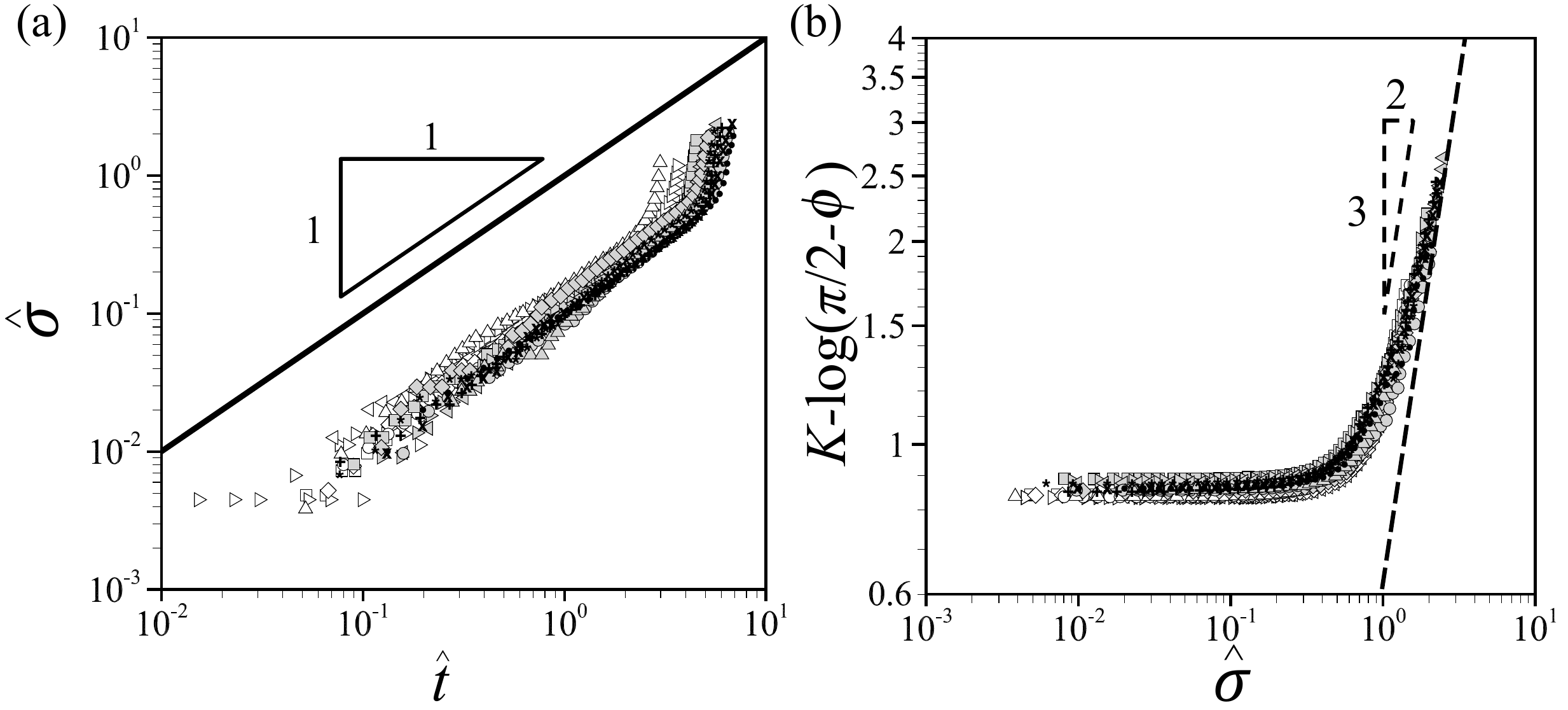}
\caption{Large deflection results from figure~\ref{large-results}
plotted using the normalised
  variables defined in (\ref{ldr:nondim}).
  (a)~$\hat{\sigma}$ versus $\hat{t}$;
the power law $\hat{\sigma}=\hat{t}$ predicted by the large time
asymptotic analysis is indicated using a solid line.
(b)~$K-\log\left(\pi/2-\phi\right)$ versus $\hat{\sigma}$, where
$K$ is defined by (\ref{Klog}); the predicted behaviour
(\ref{Klogsim}) is indicated by the dashed curve.}
  \label{large-results-scaling}
\end{figure}

In conclusion, when the beam sags to a nearly vertical configuration,
we predict that the liquid front should grow linearly with time, i.e.\
$\hat{\sigma}\left(\hat{t}\right)\sim\hat{t}$,
and that the deflection angle and normalised free boundary position
should satisfy the relation
\begin{equation}\label{ldr:phat}
\phi\sim
\frac{\pi}{2}-\frac{8\sqrt{\pi}\left(\sqrt{2}-1\right) 
{\hat{\sigma}}^{1/4}\mathrm{e}^{-2{\hat{\sigma}}^{3/2}/3}}
{3^{1/6}\Gamma(2/3)}.
\end{equation}
In figure~\ref{large-results-scaling}(a), we re-plot the large
deflection results for $\sigma$ from figure~\ref{large-results}(b)
using the normalised variables defined in equation (\ref{ldr:nondim}).
We find that the data collapse onto a single curve, which agrees quite
well with the linear behaviour predicted by equation
(\ref{ldr:nvAsi}), although with an $O(1)$ disagreement in the
prefactor. We discuss this disparity further in \S\ref{sec:concs}.
To test the predicted relation (\ref{ldr:phat}), in
figure~\ref{large-results-scaling}(b) we plot $K-\log(\pi/2-\phi)$
versus $\hat{\sigma}$, where $K$ is used as shorthand for the constant
\begin{equation}\label{Klog}
K=\log\left(\frac{8\sqrt{\pi}
\left(\sqrt{2}-1\right)}{3^{1/6}\Gamma(2/3)}\right)\approx1.284.
\end{equation}
Again we observe a dramatic collapse of the data in
figure~\ref{large-results-scaling}(b), as well as
approximate convergence towards the asymptotic behaviour
\begin{equation}\label{Klogsim}
K-\log\left(\frac{\pi}{2}-\phi\right)\sim
\frac{2}{3}\,{\hat{\sigma}}^{3/2}-\frac{1}{4}\,\log\hat{\sigma}
\end{equation}
corresponding to (\ref{ldr:phat}), which is indicated by a dashed
curve.

\section{Discussion and conclusions}
\label{sec:concs}

We have studied both experimentally and theoretically the flow of a
thin liquid rivulet along a flexible beam that is fixed at one
end. The propagation of the liquid and the deflection of the beam are
intimately coupled: the weight of the liquid causes the beam to bend
which, in turn, determines the effective body force driving the
spreading of the liquid.
Thus, this problem naturally combines two classic nonlinear mechanics
problems in fluid mechanics and elasticity.

In analysing the problem mathematically, two distinct limits for the beam deflection were identified.
In the ``small deflection'' limit, the contributions to the liquid flux from the slope of the beam and of the free surface are comparable, but the beam equations may be linearised.
In the ``large deflection'' limit, the full nonlinear beam equations must be solved, but the liquid flux is dominated by the large beam slope. In either case, the mathematical model may be simplified and then made parameter free by a suitable normalisation.
We demonstrated that the scalings thus predicted by the theory provide a very good collapse of a wide range of experimental data.

We found three distinct limiting solutions to
the mathematical models obtained in the small- and large-deflection limits. The resulting power law solutions for
the position of the liquid front and the beam deflection
are collected in table~\ref{PowerLaws}.
The ``small time'' solution is valid while the beam deflection is so
small as to have a negligible influence on the liquid, which therefore
spreads as if on a horizontal substrate.
The ``intermediate time'' solution occurs when the beam deflection is large enough to dominate the spreading of the liquid, but still small enough for the beam equations to be linearised.
Finally, the ``large time'' solution emerges when the liquid has spread so far as to weigh the beam down almost to the vertical.

By comparison with the time-scales used to normalise the problem in equations (\ref{sdr:ndvars}) and (\ref{ldr:nondim}), we infer that the corresponding ranges for the dimensionless time $t$ are given by
\begin{subequations}
\begin{align}
\text{small time:} & & &
t\ll 
t_\text{small}=\left(\frac{\mu^4\gamma E^5b^{10}w^{15}}{\rho^{10}g^{10}q^{12}}\right)^{1/16},
\\\label{inttineq}
\text{intermediate time:} & & &
t_\text{small}
\ll t \ll 
t_\text{large},
\\
\text{large time:} & & &
t \gg
t_\text{large}=
\left(\frac{\mu^2w^9b^7E^3}{\rho^5g^5q^7}\right)^{1/9}. 
\end{align}
\end{subequations}
The intermediate time regime can exist only if the lower bound in
(\ref{inttineq}) is significantly smaller than the upper bound. The
dimensionless ratio of the two time-scales is given by
\begin{equation}
\frac{t_\text{small}}{t_\text{large}}=
\frac{\gamma^{1/16}\mu^{1/36}q^{1/36}}
{\rho^{5/72}g^{5/72}b^{11/72}w^{1/16}E^{1/48}}.
\end{equation}
For the experimental parameter values, we find that
$t_\text{small}/t_\text{large}$ is in the range 0.6--0.85, that is,
smaller than one but not very small. This perhaps helps to explain why
the intermediate regime appears to persist only briefly in
Figure~\ref{scale_results}.

\begin{table}
\centering
\begin{tabular}{cccc}
& ~~Small time~~ & ~~Intermediate time~~ & ~~Large time~~
\\  \hline 
Rivulet length $\sigma$ & $\propto t^{4/5}$
& $\propto t^4$ & $\propto t$
\\\hline 
Beam deflection $\phi$ & $\propto t^{4}$ & $\propto t^{9}$
& $\simeq\pi/2$
\\\hline
\end{tabular}
\caption{Asymptotic solutions for liquid propagation along a flexible beam and the deformation of the beam. In this table, $\sigma$ and $\phi$ represent the length of the advancing liquid rivulet and the deflection angle of the beam, respectively, and $t$ represents time.} 
\label{PowerLaws}
\end{table}

Figures~\ref{scale_results} and~\ref{large-results-scaling}
demonstrate that the power-laws listed in table~\ref{PowerLaws} agree
quite well with experimental results. However, there is some
discrepancy in the pre-factors.
This is probably due to the simplified constitutive relations
(\ref{crPQsimp}) for the liquid pressure and flux used in our
mathematical analysis.
The dramatic collapse of the experimental data and the apparent agreement with the predicted power law exponents both support our claim that the relations (\ref{crPQsimp}) contain the relevant physics and exhibit the right qualitative behaviour.
However,
as pointed out in \S\ref{sec:geq}, these relations are strictly valid
only if the Bond number $\Bo$ and the ratio $A/b^2$ are both small,
neither of which is universally true in the experiments.

If the Bond number is not assumed to be small, then, under the
lubrication approximation,
the free surface of the liquid layer satisfies the Young--Laplace equation, balancing the capillary and hydrostatic pressures.
Provided $A/b^2$ is small,
the relation between the base pressure $P$ and the cross-sectional area $A$ may then in principle be expressed in terms of hyperbolic functions \citep[as in][]{paterson2013pinning}.
On the other hand, if $A/b^2$ is not small, implying that the liquid layer is not thin, then in general the flux $Q$ can only be found numerically, by solving Poisson's equation for the liquid velocity along the beam.
In principle, one can address each of these mathematical complications
in a full computational solution of the problem, but it would seem to
preclude any possibility of finding universal analytical predictions
like those listed in table~\ref{PowerLaws}.

As shown in Appendix~\ref{ss:deriv}, one can relatively easily
calculate the first corrections to
the leading-order constitutive relations (\ref{crPQsimp})
when $A/b^2$ and $\Bo$ are small but nonzero, namely
\begin{subequations}
\begin{align}\label{coP}
P(s,t)&\sim\left(\frac{3\gamma}{2b^3}\right)A
\left[1-\frac{27}{40}\,\frac{A^2}{b^4}+\frac{2}{5}\Bo\cos\psi+\cdots
\right],
\\\label{coQ}
Q(s,t)&\sim\frac{9A^3}{70\mu b^2}\left(\rho g\sin\psi-
\frac{\partial P}{\partial s}\right)
\left[1-\frac{2}{5}\,\frac{A^2}{b^4}-\frac{1}{45}\Bo\cos\psi
+\cdots\right]. 
\end{align}
\end{subequations}
In the small-time regime where $\psi\rightarrow0$, we therefore find
that
\begin{equation}
Q(s,t)\sim-\frac{27\gamma}{140\mu b^5}\,
A^3\frac{\partial A}{\partial s}\left[
1+\frac{17}{45}\Bo-\frac{97}{40}\,\frac{A^2}{b^4}\right].
\end{equation}
Thus,
inclusion of the transverse gravitational term proportional to
$\Bo$ increases the spreading rate, while the geometric correction
proportional to $A^2/b^4$ decreases the spreading rate.
It is conceivable that the combination of these effects could help to
explain the discrepancy observed in figure~\ref{scale_results}(a),
where the theory appears consistently to over-predict the spreading
rate by a factor of 2--3. In the large-time regime where
$\psi\rightarrow\pi/2$ and the pressure gradient becomes negligible,
we instead have
\begin{equation}
Q(s,t)\sim\frac{9\rho g}{70\mu b^2}\,A^3
\left[1-\frac{2}{5}\,\frac{A^2}{b^4}\right].
\end{equation}
The leading-order term is equivalent to equation (1) of
\cite{wilson2005uni}, and we observe that the geometric correction
always decreases the spreading rate.
This result is consistent with the observation in
figure~\ref{large-results-scaling}(a) that the simplified theory
persistently over-predicts the spreading rate, by a factor of around
5--10.

Finally, we note that the 
wettability of the substrate to the working fluid appears to give rise
to a rather large advancing contact angle.
In figure~\ref{expsetup}, for example, we observe a blunt free surface
profile and
the formation of a noticeable bulge near the advancing front of the
liquid film. Our simplified thin-film model is unlikely to capture
accurately the quantitative behaviour of this localised structure.
It may be that capillary effects near the advancing contact line limit
the propagation of the front such that it lags behind the
spreading rate of the thin film, resulting in accumulation of liquid
into the observed bulge near the front.

\vspace{5mm}

We are very grateful to an anonymous referee, whose insightful
suggestions resulted in significant improvements to this paper.

\appendix

\section{Derivation of constitutive relations}
\label{ss:deriv}


Here we sketch the derivation of the constitutive relations
(\ref{crPQsimp}) for the base pressure~$P$ and the flux $Q$ in the
rivulet. A schematic of the cross-section of the rivulet is shown in
figure~\ref{schematic}(b).
The $y$- and $\hat{z}$-axes are parallel and normal respectively to
the upper surface of the beam, which is at $\hat{z}=0$.
Note the distinction between $\hat{z}$ and the vertical coordinate~$z$
defined in figure~\ref{schematic}(a); they are related by
\begin{equation}
z=-\int_0^s\sin\psi\,\mathrm{d}s+\hat{z}\cos\psi.
\end{equation}
The free surface is denoted by $\hat{z}=h(y)$, where the parametric
dependence upon time $t$ and arc-length $s$ along the beam has been
temporarily suppressed.

Under the assumptions of lubrication theory, the pressure in the
rivulet is purely hydrostatic, and the free surface profile $h(y)$
satisfies the Young--Laplace equation
\begin{equation}\label{app:ly}
\frac{\gamma h''(y)}{\left[1+h'(y)^2\right]^{3/2}}
=\rho gh(y)\cos\psi-P.
\end{equation}
The solution of (\ref{app:ly}) subject to
$h'(0)=h(b)=0$ determines $h(y)$ and hence
\begin{equation}
A=\int_{-b}^b h(y)\,\mathrm{d}y
\end{equation}
in terms of $P$ and $\psi$; inversion of this relation then in
principle gives $P$ as a function of $A$ and $\psi$.

The velocity $u$ in the $s$-direction satisfies Poisson's equation
in the form
\begin{equation}\label{app:pois}
\mu\left(\frac{\partial^2u}{\partial y^2}
+\frac{\partial^2u}{\partial{\hat{z}}^2}\right)=
\frac{\partial P}{\partial s}-\rho g\sin\phi.
\end{equation}
The imposition of zero slip at the base and a zero shear stress at the
free surface leads to the boundary conditions
\begin{align}\label{app:ubcs}
u=0\quad\text{at }\hat{z}=0,
& & &
\frac{\partial u}{\partial\hat{z}}-h'(y)\frac{\partial u}{\partial y}
=0\quad\text{at }\hat{z}=h(y).
\end{align}
The solution of (\ref{app:pois}) subject to (\ref{app:ubcs}) in
principle determines $u$ and hence
\begin{equation}
Q=\int_{-b}^b\int_0^{h(y)}u\left(y,\hat{z}\right)
\,\mathrm{d}\hat{z}\,\mathrm{d}y
\end{equation}
in terms of $A$, $P$ and $\psi$.

To obtain the simplified expressions (\ref{crPQsimp}), we assume that
the rivulet is thin and that gravity is subdominant to surface
tension, so that the cross-sectional Bond number is small.
We formalize these assumptions by non-dimensionalising the above
equations and boundary conditions as follows:
\begin{align}
y&=b\tilde{y},
&
\left\{\hat{z},h\right\}&=\epsilon b\left\{\tilde{z},\tilde{h}\right\},
&
P&=\left(\frac{\epsilon\gamma}{b}\right)\tilde{P},
&
u&=\frac{\epsilon^2 b^2}{\mu}\left(\rho g\sin\phi-
\frac{\partial P}{\partial s}\right)\tilde{u},
\end{align}
where $\epsilon\rightarrow0$ in the limit of a thin rivulet.
Henceforth the tildes will be dropped to reduce clutter.
We also define
\begin{equation}
\Bo\cos\phi=\epsilon^2\beta
\end{equation}
and suppose that $\beta=O(1)$ as $\epsilon\rightarrow0$:
this conveniently ensures that gravitational and geometric corrections
enter at the same order.

The Young-Laplace equation (\ref{app:ly}) becomes
\begin{equation}\label{ylhd}
\frac{h''(y)}{\left[1+\epsilon^2h'(y)^2\right]^{3/2}}=
\epsilon^2\beta h(y)-P,
\end{equation}
which is subject to $h(\pm1)=0$.
The cross-sectional area is then given by
\begin{equation}\label{Adb2}
\frac{A}{\epsilon b^2}=2\int_0^1 h(y)\,\mathrm{d}y.
\end{equation}
We then write $h$ and $P$ as asymptotic expansions in powers of
$\epsilon^2$, i.e.
\begin{align}
h(y)&\sim h_0(y)+\epsilon^2 h_1(y)+\cdots,
&
P&\sim P_0+\epsilon^2P_1+\cdots.
\end{align}
Equation (\ref{ylhd}) may be solved successively for $h_0$, $h_1$,
\ldots, and then the condition (\ref{Adb2}) determines $P_0$,
$P_1$,\ldots.
After halting this procedure at order $\epsilon^2$ and returning to
dimensional variables, we find the approximation (\ref{coP}) for $P$.
The first term corresponds to the model (\ref{crPQsimp}) used in
the body of the paper. The following two terms are the first
corrections arising from the nonlinear geometry and from gravity,
respectively.

Next we solve for the normalised velocity $u(y,z)$, which satisfies
the problem
\begin{subequations}
\begin{gather}
\frac{\partial^2u}{\partial z^2}
+\epsilon^2\frac{\partial^2u}{\partial y^2}=-1,
\\
u=0 \quad\text{at }z=0,\qquad\qquad
\frac{\partial u}{\partial z}=
\epsilon^2h'(y)\frac{\partial u}{\partial y}
\quad\text{at }z=h(y).
\end{gather}
\end{subequations}
As above,
we solve by writing $u$ as an asymptotic expansion in powers of
$\epsilon^2$, and the normalized flux is then given by
\begin{equation}
Q=2\int_0^1\int_0^{h(y)}u(y,z)\,\mathrm{d}z\,\mathrm{d}y.
\end{equation}
We truncate the expansion at $O(\epsilon^2)$ and return to
dimensional variables to obtain the approximation (\ref{coQ}) for
$Q$.
Again the leading term gives the model (\ref{crPQsimp}), and the
subsequent terms give the first corrections in $A/b^2$ and $\Bo$.

\section{Solution of numerical shooting problems}

\subsection{Small deflection, small $\tilde{t}$}
\label{ss:sdst}

We have to solve the ODE (\ref{t0f}) subject to the boundary 
conditions (\ref{t0bcs}). 
We first make the problem autonomous via the transformation 
\begin{align}
\eta&=c\mathrm{e}^{-\xi},
&
f(\eta)&=\eta^{2/3}F(\xi),
\end{align}
so that $F(\xi)$ satisfies the ODE 
\begin{equation}
F''+\frac{3{F'}^2}{F}-\frac{4F'}{5F^3}
-\frac{13F'}{3}+\frac{10F}{9}+\frac{1}{3F^2}=0 
\end{equation}
and the initial conditions 
\begin{equation}
F(\xi)\rightarrow0,
\quad 
F'(\xi)F(\xi)^3\rightarrow0 
\quad\text{as }\xi\rightarrow0. 
\end{equation}
There is a unique solution of this initial-value problem, with the 
asymptotic behaviour 
\begin{equation}
F(\xi)\sim\left(\frac{12\xi}{5}\right)^{1/3}\left\{
1+\frac{47\xi}{96}+\frac{8983\xi^2}{64512}+\cdots 
\right\}
\quad\text{as }\xi\rightarrow0. 
\end{equation}
We use this behaviour to integrate from a small positive value of 
$\xi$. The initial condition $f(0)^3f'(0)=-1$ then allows us to 
determine both $c$ and the value of $f(0)$ from the far-field 
behaviour of $F(\xi)$, using 
\begin{equation}
\mathrm{e}^{-2\xi/3}F(\xi)\rightarrow c^{-2/3}f(0),
\quad 
\mathrm{e}^{-5\xi/3}F(\xi)^3 
\left(F'(\xi)-\frac{2}{3}\,F(\xi)\right)\rightarrow 
c^{-5/3}
\quad\text{as }\xi\rightarrow\infty. 
\end{equation}
We thus obtain the values 
$f(0)\approx1.32628$ and $c\approx1.00101$. 
The normalised deflection angle $g(\eta)$ is then determined by the 
integral (\ref{t0gint}), from which we find that 
\mbox{$g(0)\approx0.129117$}.

The numerical solutions thus obtained for the functions $f(\eta)$ and 
$g(\eta)$ are plotted in figure~\ref{fig:smalltfg}. 

\subsection{Small deflection, large $\tilde{t}$}
\label{ss:sdlt}

The small-deflection, large-$\tilde{t}$ problem from \S\ref{ss:sdr}
leads to the system of ODEs (\ref{tinffgde}) 
and boundary conditions (\ref{tinffgbcs}), (\ref{tinfflux}) 
for the similarity solution variables $f(\eta)$ and $g(\eta)$. 
We now make the problem autonomous by defining 
\begin{align}
\eta&=c\mathrm{e}^{-\xi},
&
f(\eta)&=\eta^{-2/3}F(\xi),
&
g(\eta)&=\eta^{7/3}G(\xi),
\end{align}
so that $F$ and $G$ satisfy the ODEs 
\begin{align}
F'&=\frac{F\left(1+3F^2G'-F^2G\right)}{3\left(4-3F^2G\right)},
&
G'''-4G''+\frac{13}{3}G'-\frac{28}{27}G+F&=0,
\end{align}
and boundary conditions 
\begin{align}
G'(0)&=\frac{7}{3}G(0),
&
G''(0)&=\frac{49}{9}G(0),
&
F(0)&=2G(0)^{-1/2}. 
\end{align}
The conditions (\ref{tinffgbcs}) 
at $\eta=0$ transform to the far-field conditions 
\begin{align}\label{FGxiinfty}
G(\xi)&\sim g'(0)c^{-4/3}\mathrm{e}^{4\xi/3},
&
F(\xi)&\sim g'(0)^{-1/3} c^{1/3}\mathrm{e}^{-\xi/3}
& &\text{as }\xi\rightarrow\infty. 
\end{align}
We therefore use $G(0)$ as a shooting parameter to get 
\begin{equation}
G''(\xi)-\frac{5}{3}G'(\xi)+\frac{4}{9}G(\xi)\rightarrow0 
\quad\text{as }\xi\rightarrow\infty 
\end{equation}
(corresponding to $G(\xi)\mathrm{e}^{7\xi/3}\rightarrow0$),
and then use 
(\ref{FGxiinfty}) to infer the values of $g'(0)$ and $c$. 

By following this procedure, we obtain the values 
\begin{align}
G(0)&\approx0.455938,
&
c&\approx0.0434638,
&
g'(0)&\approx0.0206883.&
\end{align}
The corresponding value of the film area and the normalised angle at 
the advancing front are then given by 
\begin{align}
f(c)&=2c^{-2/3}G(0)^{-1/2}\approx23.9603,
&
g(c)&=c^{7/3}G(0)\approx0.000302834. 
\end{align}
The resulting numerical solutions for $f(\eta)$ and $g(\eta)$ are 
plotted in figure~\ref{fig:fgvseta}.

\bibliographystyle{jfm}

\bibliography{flow_over_beam}

\begin{thebibliography}{28}
\expandafter\ifx\csname natexlab\endcsname\relax\def\natexlab#1{#1}\fi

\bibitem[Crandall {\em et~al.\/}(1978)Crandall, Lardner, Archer, Cook \&
  Dahl]{crandall1978introduction}
{\sc Crandall, S.~H., Lardner, T.~J., Archer, R.~R., Cook, N.~H. \& Dahl,
  N.~C.} 1978 {\em An Introduction to the Mechanics of Solids\/}. McGraw-Hill.

\bibitem[Davis {\em et~al.\/}(1986)Davis, Serayssol \&
  Hinch]{davis1986elastohydrodynamic}
{\sc Davis, R.~H., Serayssol, J.-M. \& Hinch, E.~J.} 1986 The
  elastohydrodynamic collision of two spheres. {\em J. Fluid Mech.\/} {\bf
  163}, 479--497.

\bibitem[Dowson \& Ehret(1999)]{dowson1999past}
{\sc Dowson, D. \& Ehret, P.} 1999 Past, present and future studies in
  elastohydrodynamics. {\em Proc. Inst. Mech. Eng. J J. Eng. Tribol.\/} {\bf
  213}~(5), 317--333.

\bibitem[Duffy \& Moffatt(1995)]{Duffy1995141}
{\sc Duffy, B.R. \& Moffatt, H.K.} 1995 Flow of a viscous trickle on a slowly
  varying incline. {\em Chem. Eng. J. Bioch. Eng.\/} {\bf 60}~(1-3), 141 --
  146.

\bibitem[Duffy \& Moffatt(1997)]{EJM:43785}
{\sc Duffy, B.~R. \& Moffatt, H.~K.} 1997 A similarity solution for viscous
  source flow on a vertical plane. {\em Eur. J. of Appl. Math.\/} {\bf 8},
  37--47.

\bibitem[Flitton \& King(2004)]{flitton2004moving}
{\sc Flitton, J.~C. \& King, J.~R.} 2004 Moving-boundary and fixed-domain
  problems for a sixth-order thin-film equation. {\em Eur. J. Appl. Math.\/}
  {\bf 15}~(06), 713--754.

\bibitem[Fritz {\em et~al.\/}(2013)Fritz, Seminara, Roper, Pringle \&
  Brenner]{fritz2013natural}
{\sc Fritz, J.~A., Seminara, A., Roper, M., Pringle, A. \& Brenner, M.~P.} 2013
  A natural o-ring optimizes the dispersal of fungal spores. {\em J. Roy. Soc.
  Interface\/} {\bf 10}~(85), 20130187.

\bibitem[Gart {\em et~al.\/}(2015)Gart, Mates, Megaridis \&
  Jung]{gart2015droplet}
{\sc Gart, S., Mates, J.~E., Megaridis, C.~M. \& Jung, S.} 2015 Droplet
  impacting a cantilever: A leaf-raindrop system. {\em Phys. Rev. Appl.\/} {\bf
  3}~(4), 044019.

\bibitem[Gilet \& Bourouiba(2015)]{gilet2015fluid}
{\sc Gilet, T. \& Bourouiba, L.} 2015 Fluid fragmentation shapes rain-induced
  foliar disease transmission. {\em J. Roy. Soc. Interface\/} {\bf 12}~(104),
  20141092.

\bibitem[Gohar(2001)]{gohar2001elastohydrodynamics}
{\sc Gohar, R.} 2001 {\em Elastohydrodynamics\/}. World Scientific.

\bibitem[Hewitt {\em et~al.\/}(2015)Hewitt, Balmforth \&
  De~Bruyn]{hewitt2015elastic}
{\sc Hewitt, I.~J., Balmforth, N.~J. \& De~Bruyn, J.~R.} 2015 Elastic-plated
  gravity currents. {\em Eur. J. Appl. Math.\/} {\bf 26}~(01), 1--31.

\bibitem[Howell {\em et~al.\/}(2013)Howell, Robinson \&
  Stone]{howell2013gravity}
{\sc Howell, P.~D., Robinson, J. \& Stone, H.~A.} 2013 Gravity-driven thin-film
  flow on a flexible substrate. {\em J. Fluid Mech.\/} {\bf 732}, 190--213.

\bibitem[Huppert(1982{\natexlab{{\em a\/}}})]{huppert1982flow}
{\sc Huppert, H.~E.} 1982{\natexlab{{\em a\/}}} Flow and instability of a
  viscous current down a slope. {\em Nature\/} {\bf 300}~(5891), 427--429.

\bibitem[Huppert(1982{\natexlab{{\em b\/}}})]{huppert1982propagation}
{\sc Huppert, H.~E.} 1982{\natexlab{{\em b\/}}} The propagation of
  two-dimensional and axisymmetric viscous gravity currents over a rigid
  horizontal surface. {\em J. Fluid Mech.\/} {\bf 121}, 43--58.

\bibitem[Leslie {\em et~al.\/}(2013)Leslie, Wilson \& Duffy]{FLM:8825124}
{\sc Leslie, G.~A., Wilson, S.~K. \& Duffy, B.~R.} 2013 Three-dimensional
  coating and rimming flow: a ring of fluid on a rotating horizontal cylinder.
  {\em J. Fluid Mech.\/} {\bf 716}, 51--82.

\bibitem[Lister {\em et~al.\/}(2013)Lister, Peng \& Neufeld]{lister2013viscous}
{\sc Lister, J.~R., Peng, G.~G. \& Neufeld, J.~A.} 2013 Viscous control of
  peeling an elastic sheet by bending and pulling. {\em Phys. Rev. Lett.\/}
  {\bf 111}~(15), 154501.

\bibitem[Mow {\em et~al.\/}(1992)Mow, Ratcliffe \& Poole]{mow1992cartilage}
{\sc Mow, V.~C., Ratcliffe, A. \& Poole, A.~R.} 1992 Cartilage and diarthrodial
  joints as paradigms for hierarchical materials and structures. {\em
  Biomaterials\/} {\bf 13}~(2), 67--97.

\bibitem[Pang {\em et~al.\/}(2014)Pang, Kim, Liu \& Stone]{pang2014soft}
{\sc Pang, Y., Kim, H., Liu, Z. \& Stone, H.~A.} 2014 A soft microchannel
  decreases polydispersity of droplet generation. {\em Lab Chip\/} {\bf
  14}~(20), 4029--4034.

\bibitem[Paterson {\em et~al.\/}(2013)Paterson, Wilson \&
  Duffy]{paterson2013pinning}
{\sc Paterson, C., Wilson, S.~K. \& Duffy, B.~R.} 2013 Pinning, de-pinning and
  re-pinning of a slowly varying rivulet. {\em Eur. J. Mech. B-Fluid.\/} {\bf
  41}, 94--108.

\bibitem[Salez \& Mahadevan(2015)]{salez2015elastohydrodynamics}
{\sc Salez, T. \& Mahadevan, L.} 2015 Elastohydrodynamics of a sliding,
  spinning and sedimenting cylinder near a soft wall. {\em J. Fluid Mech.\/}
  {\bf 779}, 181--196.

\bibitem[Sekimoto \& Leibler(1993)]{sekimoto1993mechanism}
{\sc Sekimoto, K. \& Leibler, L.} 1993 A mechanism for shear thickening of
  polymer-bearing surfaces: elasto-hydrodynamic coupling. {\em Europhys.
  Lett.\/} {\bf 23}~(2), 113.

\bibitem[Shelley \& Zhang(2011)]{shelley2011flapping}
{\sc Shelley, M.~J. \& Zhang, J.} 2011 Flapping and bending bodies interacting
  with fluid flows. {\em Ann. Rev. Fluid Mech.\/} {\bf 43}, 449--465.

\bibitem[Skotheim \& Mahadevan(2005)]{skotheim2005soft}
{\sc Skotheim, J.~M. \& Mahadevan, L.} 2005 Soft lubrication: the
  elastohydrodynamics of nonconforming and conforming contacts. {\em Phys.
  Fluids\/} {\bf 17}~(9), 092101.

\bibitem[Tony {\em et~al.\/}(2006)Tony, Lauga \& Hosoi]{tony2006experimental}
{\sc Tony, S.~Y., Lauga, E. \& Hosoi, A.~E.} 2006 Experimental investigations
  of elastic tail propulsion at low reynolds number. {\em Phys. Fluids\/} {\bf
  18}~(9), 091701.

\bibitem[Wexler {\em et~al.\/}(2013)Wexler, Trinh, Berthet, Quennouz, du~Roure,
  Huppert, Lindner \& Stone]{wexler2013bending}
{\sc Wexler, J.~S., Trinh, P.~H, Berthet, H., Quennouz, N., du~Roure, O.,
  Huppert, H.~E., Lindner, A. \& Stone, H.~A.} 2013 Bending of elastic fibres
  in viscous flows: the influence of confinement. {\em J. Fluid Mech.\/} {\bf
  720}, 517--544.

\bibitem[Wiggins {\em et~al.\/}(1998)Wiggins, Riveline, Ott \&
  Goldstein]{wiggins1998trapping}
{\sc Wiggins, C.~H., Riveline, D., Ott, A. \& Goldstein, R.~E.} 1998 Trapping
  and wiggling: elastohydrodynamics of driven microfilaments. {\em Biophys.
  J.\/} {\bf 74}~(2), 1043--1060.

\bibitem[Wilson \& Duffy(2005)]{wilson2005uni}
{\sc Wilson, S.~K. \& Duffy, B.~R.} 2005 Unidirectional flow of a thin rivulet
  on a vertical substrate subject to a prescribed uniform shear stress at its
  free surface. {\em Phys. Fluids\/} {\bf 17}~(10).

\bibitem[Zheng {\em et~al.\/}(2015)Zheng, Griffiths \&
  Stone]{zheng2015propagation}
{\sc Zheng, Z., Griffiths, I.~M. \& Stone, H.~A.} 2015 Propagation of a viscous
  thin film over an elastic membrane. {\em J. Fluid Mech.\/} {\bf 784},
  443--464.

\end{thebibliography}

\end{document}